\documentclass[aps,prb,reprint,superscriptaddress]{revtex4-2}
\usepackage{graphicx}
\usepackage{latexsym}
\usepackage{amsmath}
\usepackage{amssymb}
\usepackage{amsthm}
\usepackage{amsfonts}
\usepackage{color}
\usepackage{xcolor}
\usepackage{verbatim}
\usepackage{url}
\usepackage{array}
\usepackage{multirow}
\usepackage{algorithm}
\usepackage{algorithmicx}
\usepackage{algpseudocode}
\usepackage{txfonts}
\usepackage{physics}
\usepackage{siunitx}
\usepackage{hyperref}
\usepackage{caption}
\usepackage{subcaption}
\usepackage{graphicx}
\hypersetup{
    colorlinks,
    linkcolor={red!50!black},
    citecolor={red!70!black},
    urlcolor={red!80!black}
}
\newif\ifmarked
\markedfalse
\definecolor{revblue}{RGB}{0,70,220}
\newcommand{\rev}[1]{\ifmarked{\color{revblue}#1}\else#1\fi}

\newcommand{\T}[1]{#1^{\intercal}}
\newcommand{\ERI}[2]{\braket{#1}{#2}}
\newcommand{\dbERI}[2]{\mel{#1}{}{#2}}
\newcommand{\ee}{\text{ee}}
\newcommand{\eh}{\text{eh}}
\newcommand{\he}{\text{he}}
\newcommand{\hh}{\text{hh}}

\newcommand{\bO}{\boldsymbol{0}}
\newcommand{\bI}{\boldsymbol{1}}
\newcommand{\bH}{\boldsymbol{H}}
\newcommand{\bF}{\boldsymbol{F}}

\newcommand{\bv}{\boldsymbol{v}}
\newcommand{\bV}{\boldsymbol{V}}
\newcommand{\bU}{\boldsymbol{U}}
\newcommand{\bW}{\boldsymbol{W}}
\newcommand{\bD}{\boldsymbol{D}}
\newcommand{\bR}{\boldsymbol{R}}
\newcommand{\bG}{\boldsymbol{G}}
\newcommand{\br}{\boldsymbol{r}}
\newcommand{\bx}{\boldsymbol{x}}
\newcommand{\ba}{\boldsymbol{a}}
\newcommand{\balpha}{\boldsymbol{\alpha}}
\newcommand{\bkappa}{\boldsymbol{\kappa}}
\newcommand{\bDelta}{\boldsymbol{\Delta}}
\newcommand{\eps}{\epsilon}
\newcommand{\beps}{\boldsymbol{\epsilon}}
\newcommand{\bSig}{\boldsymbol{\Sigma}}
\newcommand{\bchi}{\boldsymbol{\chi}}

\newcommand{\hH}{\Hat{H}}
\newcommand{\upar}{\uparrow}
\newcommand{\dwar}{\downarrow}
\newcommand{\HF}{\text{HF}}
\newcommand{\HFB}{\text{HFB}}
\newcommand{\BCS}{\text{BCS}}
\newcommand{\GM}{\text{GM}}

\newcommand{\ii}{\text{i}}

\newcommand{\anh}[2]{{\hat{#1}^{\vphantom{\dagger}}}_{#2}}
\newcommand{\cre}[2]{{\hat{#1}^{\dagger}}_{#2}}

\graphicspath{{./Figures/}{./}}

\newcommand{\LCPQ}{Laboratoire de Chimie et Physique Quantiques (UMR 5626), Universit\'e de Toulouse, CNRS, Toulouse, France}
\newcommand{\UnHam}{Department of Chemistry, University of Hamburg, 22761 Hamburg, Germany; The Hamburg Centre for Ultrafast Imaging (CUI), Hamburg 22761, Germany}

\begin{document}

\title{Bogoliubov self-consistent $GW$ and second-order Green's function methods\\ for attractive fermionic interactions}

\author{Mauricio Rodr{\'i}guez-Mayorga}
	\email{marm3.14@gmail.com}
	\affiliation{\LCPQ}

\author{Johannes \surname{T\"olle}}
        \email{johannes.toelle@uni-hamburg.de}
        \affiliation{\UnHam}

\author{Pierre-Fran\c{c}ois Loos}
	\email{loos@irsamc.ups-tlse.fr}
	\affiliation{\LCPQ}

\begin{abstract}
Describing fermionic systems with attractive interactions, such as superconductors and atomic nuclei, beyond the mean-field level remains a central challenge. In this work, we investigate the performance of Bogoliubov fully self-consistent $GW$ (Bog.~sc$GW$) and second-order Green's function (Bog.~scGF2) methods, implemented within a unified imaginary-time and imaginary-frequency framework. Using \rev{two attractive Hubbard lattices ($4\times4$ and $2\times8$) as a model study and benchmarking} against full configuration interaction (FCI), we find that\rev{, for these systems,} Bog.~scGF2 provides a more accurate description than Bog.~sc$GW$.
In particular, Bog.~sc$GW$ systematically overestimates the correlation energy, with errors increasing rapidly as the attractive interaction strength grows. These results \rev{suggest} the importance of exchange contributions in \rev{such} strongly correlated fermionic systems, even when exchange effects are included only at low order in perturbation theory.
\end{abstract}
\maketitle

\section{Introduction}

Attractive interactions between fermions give rise to pairing, one of the most striking collective quantum phenomena in nature. 
It underlies superconductivity, discovered in 1911 by H.~K.~Onnes~\cite{levelt2002fluids} and later extended to high-temperature materials~\cite{bednorz1986possible}, but also the pairing of nucleons in atomic nuclei, which manifests itself as an energy gap in the intrinsic excitation spectrum~\cite{bohr1958possible}, and the superfluidity of nuclear matter and neutron stars~\cite{dean2003pairing}. 
Understanding and predicting the properties of such paired fermionic systems from first principles, therefore, remains a central objective of both electronic structure theory and nuclear many-body theory~\cite{bender2003self,hergert2020guided}.

In 1957, Bardeen, Cooper, and Schrieffer (BCS) proposed a microscopic theory of conventional superconductivity~\cite{bardeen1957theory}, in which an effective attraction between electrons near the Fermi surface leads to correlated pairs (Cooper pairs) and to an energy gap in the excitation spectrum. 
The same pairing mechanism applies to nucleons in nuclei~\cite{bohr1958possible,dean2003pairing}. 
More generally, BCS-type pairing arises whenever an effective attraction between fermions overcomes the repulsive contributions near the Fermi surface. 
In conventional superconductors, this attraction is mediated by phonons and can be obtained, for example, from the electron-phonon Hamiltonian via a Fröhlich canonical transformation~\cite{kolley1990derivation}, whereas in nuclei it originates from the underlying nucleon-nucleon interaction~\cite{dean2003pairing}. 

In accordance with BCS theory~\cite{bardeen1957theory}, including an attractive interaction between fermions, the wavefunction of the system can be approximated as a BCS state~\cite{bardeen1957theory,valatin1958comments}
\begin{equation} \label{eq:bcs}
     \ket{\Phi_{\BCS}} = \prod_p \qty( \frac{1+g_p \cre{a}{p_\upar}\cre{a}{p_\dwar} }{\sqrt{1+\abs*{g_p}^2}} ) \ket{0}
\end{equation}
where $\cre{a}{p_\upar}$ ($\anh{a}{p_\dwar}$) is the usual fermionic creation (annihilation) operators, $p_\upar$ and $p_\dwar$ refer to the spin-up and spin-down single-particle states related by time-reversal symmetry, $g_p$ is a normalization constant, and $\ket{0}$ corresponds to the (fermionic) vacuum state. 

Introducing the fermionic Bogoliubov transformation~\cite{bogoliubov1958new} 
\begin{equation} \label{eq:bogoliubov_trans}
	\begin{pmatrix}
		\anh{\balpha}{}  \\
		\cre{\balpha}{}
	\end{pmatrix} 
	=
	\begin{pmatrix}
		\bU^\dagger & \bV^\dagger \\
		\T{\bV} & \T{\bU}
	\end{pmatrix} 
	\cdot
	\begin{pmatrix}
		\anh{\ba}{}  \\
		\cre{\ba}{}
	\end{pmatrix} 
	= \bW^\dagger \cdot
	\begin{pmatrix}
		\anh{\ba}{}  \\
		\cre{\ba}{}
	\end{pmatrix} 
\end{equation}
with $\bW$ being a unitary matrix (and the definition of $\bU$ and $\bV$ given in Sec.~\ref{subsec:hfb}), one can define a new set of fermionic operators $\anh{\balpha}{}$ ($\cre{\balpha}{}$) representing the annihilation (creation) of some fermionic quasiparticles formed as a combination of particles and holes of opposite spin~\cite{ring2004nuclear,valatin1958comments}. 

Expressing in terms of Bogoliubov quasiparticle operators, the BCS state corresponds to the quasiparticle vacuum, i.e., 
\begin{align}
    \label{eq:vacuum_qp}
	\ket{\Phi_{\BCS}} & = \prod_p \anh{\alpha}{p} \ket{0}
	& 
	\anh{\alpha}{q} \ket{\Phi_{\BCS}} & = 0 \; \forall q
\end{align}	
Consequently, the BCS state is a single-reference state (in the sense that it is defined as a vacuum with respect to a set of quasiparticle operators). Describing the system at this level, therefore, amounts to adopting a mean-field approximation. The corresponding self-consistent equations are obtained from a generalization of the Hartree--Fock method known as the Hartree--Fock--Bogoliubov (HFB) formalism~\cite{ring2004nuclear}, which is briefly reviewed in Sec.~\ref{subsec:hfb}. HFB is the workhorse for treating pairing correlations in nuclear structure, where it underlies the widely used self-consistent mean-field models based on Skyrme, Gogny, and relativistic energy functionals~\cite{bender2003self}.
As a mean-field theory, HFB neglects dynamical correlation effects that are essential for an accurate description of strongly interacting fermionic systems, from the electronic structure of correlated materials to atomic nuclei~\cite{ring2004nuclear,sokolov2015transformed,signoracci2014ab,qiu2019particle,hergert2020guided}. Going beyond the BCS/HFB mean-field picture is thus a challenge shared by all these fields. 

In this context, several post-HFB approaches have been developed, in both the quantum chemistry and the nuclear physics communities, to recover the missing correlation effects, including second-order Bogoliubov Møller–Plesset perturbation theory~\cite{sokolov2015transformed}, Bogoliubov coupled-cluster methods~\cite{signoracci2014ab,signoracci2015ab,khamoshi2022agp,cui2025ab,scalesi2024impact,stolarczyk1989quasiparticles}, which have also been applied to electronic superconducting systems~\cite{cui2025ab}, Bogoliubov many-body perturbation theory~\cite{tichai2018bogoliubov}, and self-consistent Green's function approaches based on the Gorkov formalism, at second order (Bog.~scGF2)~\cite{soma2011ab,scalesi2026iconic} and beyond through the algebraic-diagrammatic construction~\cite{soma2020self,soma2020novel,stolarczyk1989quasiparticles,demartini2026algebraic}. Despite their great success, these methodologies still require significant computational resources. For example, coupled-cluster with singles and doubles (CCSD), scales as $\order*{M^6}$ with respect to the system size $M$~\cite{helgaker2013molecular}, which limits its applicability to large systems, such as the electronic superconductors to which Bogoliubov CCSD has been applied~\cite{cui2025ab}. Therefore, the search for cost-effective methods to incorporate the correlation effects beyond HFB remains an active field of research~\cite{ring2004nuclear,hergert2020guided}. 

Among the possible routes, the $GW$ approximation is particularly appealing because of its apparently good cost-effectiveness: it offers a favorable balance between accuracy and computational cost, which has contributed to its widespread adoption in both condensed-matter physics and quantum chemistry~\cite{martin2016interacting,marie2024gw}. Moreover, recent algorithmic developments have reduced the computational scaling of implementations of $GW$ to cubic with respect to system size~\cite{duchemin2021cubic,duchemin2020robust,wilhelm2021low,wilhelm2018toward,forster2021low,foorster2020low,caruso2013bond}, further enhancing its practical appeal.

Within the $GW$ approximation, electrons are dressed by dynamical correlation effects, and the resulting excitations are described in terms of quasiparticles. 
The corresponding quasiparticle energies are commonly associated with the ionization potentials (IPs) and electron affinities (EAs) of molecules, as well as band structures in extended systems.
Currently, most $GW$ calculations are performed as a one-shot correction (commonly denoted $G_0W_0$) to a mean-field starting point~\cite{martin2016interacting,reining2018gw}, which, unfortunately, introduces a dependence on the underlying mean-field reference~\cite{caruso2016benchmark}. 
To alleviate this dependence, partially self-consistent schemes have been proposed, among which the eigenvalue-only self-consistent (ev$GW$)~\cite{faleev2004all}, and quasiparticle self-consistent (qs$GW$) approaches are the most widely used~\cite{van2006quasiparticle,duchemin2025joint,kaplan2016quasi,marie2023similarity,lei2022gaussian,forster2025beyond}.

To fully remove the starting-point dependence and go beyond the quasiparticle approximation, fully self-consistent $GW$ (sc$GW$) schemes, obtained by iteratively solving Hedin's equations~\cite{hedin1965new}, have been developed and applied to a variety of molecular and solid-state systems~\cite{pokhilko2025tensor,wen2024comparing,abraham2024relativistic,yeh2022relativistic,yeh2022fully,shishkin2007self,shishkin2007accurate,caruso2016benchmark,kutepov2017ground,kutepov2017self,kutepov2020self,kutepov2021electronic}. 
Beyond quasiparticle energies and spectral functions, the sc$GW$ method provides access to total energies (for example, through the Galitskii-Migdal functional~\cite{galitskii1958application}), correlated density matrices (that can be used to evaluate physical properties, such as dipole moments)~\cite{hedin1965new}, among other quantities~\cite{pokhilko2021evaluation}. 
In the presence of pairing, the $GW$ approximation has been formulated within the Gorkov--Hedin framework~\cite{gor1958energy,essenberger2014density,linscheid2015hedin,lane2026gorkov,marie2024anomalous,marie2025anomalous}. 
Self-consistent $GW$ has already been explored in the context of superconductivity, from its combination with Eliashberg theory~\cite{davydov2018coulomb,akashi2025uniform,margine2013anisotropic} to Nambu-space calculations for the attractive Hubbard model~\cite{xiong2025application} and, very recently, for doped SrTiO$_3$~\cite{cui2026self}. 
In each case, self-consistency improves on one-shot $G_0W_0$, yet quantitative reliability for pairing remains elusive; we return to these studies in Sec.~\ref{sec:discussions} in light of our own findings. 

In the present work, we evaluate the capabilities of the fully self-consistent Bogoliubov sc$GW$ method (Bog.~sc$GW$) to incorporate dynamical correlation into the description of fermionic systems governed by attractive interactions, such as those found in superconductors and nuclei. 
We assess it against exact full configuration interaction (FCI) results for small attractive Hubbard clusters, which provide reference energies and density matrices free of statistical and imaginary-time-discretization errors. 
We focus on total energies and on the anomalous density as a measure of the effect of correlation on the paired system. 
For comparison, we also consider the Bog.~scGF2, which follows the same self-consistent procedure using a self-energy truncated at second order in the Coulomb interaction instead of the $GW$ self-energy.
The $GF2$ self-energy evaluation scales formally as $\order*{M^5}$ per imaginary-time point~\cite{neuhauser2017stochastic}, one power lower than CCSD, and, to the best of our knowledge, it has not previously been applied to electronic superconducting models. 
We further compare with MP2 and CCSD, and with their non-Bogoliubov counterparts. 
We find that Bog.~sc$GW$ performs poorly, whereas Bog.~scGF2, built with a very similar algorithm, is comparatively encouraging. 
The remainder of this paper is organized as follows. Section \ref{sec:theory} outlines the theoretical formalism, while Sec.~\ref{sec:scgw_algo} describes the computational details and algorithmic aspects of the implementation. In Sec.~\ref{sec:results}, we assess the methods on representative examples based on the attractive Hubbard model. 
Section~\ref{sec:discussions} discusses the main findings in light of previous work and perspectives, and Sec.~\ref{sec:conclusion} concludes.

\section{Theory}
\label{sec:theory} 

\subsection{The Hartree--Fock--Bogoliubov method}
\label{subsec:hfb}
To introduce the HFB method, let us first define the electronic Hamiltonian as 
\begin{equation}  
    \hH = \sum_{PQ} h_{PQ} \cre{a}{P} \anh{a}{Q} + \frac{1}{2} \sum_{PQRS} \ERI{PQ}{RS} \cre{a}{P} \cre{a}{Q} \anh{a}{S}\anh{a}{R}
\end{equation}
where the indices $P$, $Q$, $R$, and $S$ account for all the single-particle states, i.e., $\qty{ P } = \qty{ p_\upar } \cup \qty{ p_\dwar }$, the $h_{PQ}$'s contain all one-body interactions (i.e., the electronic kinetic energy and the interaction among the electrons and some external potential, e.g., the nuclei within the Born-Oppenheimer approximation) and $\ERI{PQ}{RS}$ is a two-electron integral in Dirac notation that account for the usual Coulomb repulsive interaction plus the potential phonon-mediated electron-electron attractive interaction. 

Expressing the Hamiltonian and the HFB wavefunction in the quasiparticle basis, it is straightforward to define the HFB energy as~\cite{staroverov2002optimization}
\begin{equation} \label{eq:e_hfb}
\begin{split}
    E_{\HFB}\left[ \bD,\bkappa \right] 
    & = \sum_{PQ} h_{PQ} D_{PQ} 
    \\
    & + \frac{1}{2} \sum_{PQRS} D_{PR} D_{QS} \dbERI{PQ}{RS} 
    \\
    & + \frac{1}{4} \sum_{PQRS} \kappa_{PQ}^* \kappa_{RS} \dbERI{PQ}{RS} 
\end{split}   
\end{equation}   
where $D_{PQ}=\mel{ \Phi_{\HFB} }{ \cre{a}{Q} \anh{a}{P} }{ \Phi_{\HFB} }$ are the elements of the density matrix, $\kappa_{PQ} = \mel{ \Phi_{\HFB} }{ \anh{a}{Q}  \anh{a}{P} }{  \Phi_{\HFB} }$ the so-called anomalous density matrix elements, and $\dbERI{PQ}{RS} = \ERI{PQ}{RS} - \ERI{PQ}{SR}$. 
From the anti-commutation relations of the electronic creation and annihilation operators, it can be shown that the density matrix is Hermitian (i.e., $\bD = \bD^\dagger$), while the anomalous density matrix is skew-symmetric (i.e., $\T{\bkappa} = -\bkappa$).  

Collecting the density matrix and its anomalous counterpart, one can form the generalized (or Gorkov) density matrix~\cite{gor1958energy}
\begin{equation} \label{eq:R_full}
    \bR = 
    \begin{pmatrix}
		\bD & \bkappa \\
		-\bkappa^* & \bI - \bD^*
	\end{pmatrix} 
\end{equation}
where $\bI$ is the identity matrix.
This matrix is Hermitian (i.e., $\bR^\dagger=\bR$)~\cite{ring2004nuclear} and, owing to the single-determinant character of the HFB wavefunction, it is also idempotent (i.e., $\bR^2 =\bR$). The idempotency can be exploited to express a relationship between the normal and anomalous density matrices, 
\begin{equation}
	- \bkappa \cdot \bkappa^* = \bD - \bD \cdot \bD
 \label{eq:an_den_rel}
\end{equation}
showing that the anomalous density is determined by the normal density up to an arbitrary phase factor~\cite{staroverov2002optimization}.

When spin-orbit coupling effects are not considered, as is assumed throughout this work, the generalized density matrix can be expressed in a spin-resolved form as
\begin{equation} \label{eq:R_full_so_false}
        \bR = 
        \begin{pmatrix}
        	\bD^{\upar\upar} & \bO & \bO & \bkappa^{\upar\dwar} \\
			\bO & \bD^{\dwar\dwar} & \bkappa^{\dwar\upar} & \bO   \\
			\bO & - \qty(\bkappa^{\upar\dwar})^* & \bI-\qty(\bD^{\upar\upar})^* & \bO   \\
			-\qty(\bkappa^{\dwar\upar})^* & \bO & \bO & \bI-\qty(\bD^{\dwar\dwar})^*
	\end{pmatrix}
\end{equation}
The matrix elements of the spin blocks are defined as
\begin{subequations}
\begin{align}
	D_{pq}^{\upar\upar} & = \mel{ \Phi_{\HFB} }{ \cre{a}{q_\uparrow} \anh{a}{p_\uparrow} }{ \Phi_{\HFB} }
	\\
	D_{pq}^{\dwar\dwar} & = \mel{ \Phi_{\HFB} }{ \cre{a}{q_\dwar} \anh{a}{p_\dwar} }{ \Phi_{\HFB} }
	\\
	\kappa_{pq}^{\upar\dwar} & = \mel{ \Phi_{\HFB} }{ \anh{a}{q_\dwar} \anh{a}{p_\upar} }{ \Phi_{\HFB} }
	\\
	\kappa_{pq}^{\dwar\upar} & = \mel{ \Phi_{\HFB} }{ \anh{a}{q_\upar} \anh{a}{p_\dwar} }{ \Phi_{\HFB} }
\end{align}
\end{subequations}
For spin-unpolarized calculations (i.e., $\bD^{\dwar\dwar}=\bD^{\upar\upar}$ and $\bkappa^{\dwar\upar}=-\bkappa^{\upar\dwar}$), the energy expression can be further simplified
\begin{equation}
\begin{split}
	E_{\HFB}[\bD,\bkappa] 
	& =  2 \sum_{pq} h_{p_\upar q_\upar} D^{\upar\upar}_{pq}
	\\
	& + \sum_{pqrs} D^{\upar\upar}_{pr} D^{\upar\upar}_{qs} \qty( 2 \ERI{ p_\upar q_\upar }{ r_\upar s_\upar } - \ERI{ p_\upar q_\upar }{ s_\upar r_\upar } )
	\\
	& + \sum_{pqrs} \qty(\kappa^{\upar\dwar}_{pq})^* \kappa^{\upar\dwar}_{rs} \ERI{ p_\upar q_\dwar }{ r_\upar s_\dwar } 
\label{eq:E_bcs_su}
\end{split}   
\end{equation}   
In this case, only half of the generalized density matrix elements are required to evaluate the energy. 
The spin-unpolarized generalized density matrix can therefore be written compactly as
\begin{equation}
	\bR =
	\begin{pmatrix}
		\bD^{\upar\upar} & \bkappa^{\upar\dwar} \\
		\qty(\bkappa^{\upar\dwar})^* & \bI-\qty(\bD^{\upar\upar})^*
	\end{pmatrix}
\end{equation}
From this point onward, unless otherwise stated, we omit explicit spin labels and work in a spatial orbital basis indexed by lower-case letters ($p$, $q$, $r$, $s$, \ldots). In other words, we employ a restricted HFB formalism.

To obtain the HFB ground-state wavefunction, the energy must be minimized with respect to both the density matrix and the anomalous density matrix. 
However, this optimization is subject to constraints because the normal and anomalous density matrices are not independent but are related through Eq.~\eqref{eq:an_den_rel}.
In addition, the HFB state does not preserve particle number, i.e., it breaks the so-called $U(1)$ symmetry~\cite{bardeen1957theory}. 
Therefore, when describing a system with $N$ electrons, it is customary to impose a constraint on the average particle number in the HFB state~\cite{ring2004nuclear}.
This is achieved through the Lagrangian
\begin{equation}
	\mathcal{L}[\bD,\bkappa,\mu]= E_{\HFB}[\bD,\bkappa]  + \mu \qty( N - 2\Tr\qty[\bD] )
\end{equation}
where $\mu$ is the chemical potential enforcing the particle-number constraint.
Then, taking the partial derivatives of the Lagrangian with respect to the density and the anomalous density matrices leads to the definition of the Fock operator
\begin{equation}
    F_{pq} = \frac{1}{2} \pdv{\mathcal{L}}{D_{pq}} = h_{pq} + 2J_{pq} - K_{pq} - \mu \delta_{pq}
    \label{eq:Fock}
\end{equation}
and the pairing potential
\begin{equation}
    \Delta_{pq} = \frac{1}{2} \pdv{\mathcal{L}}{\kappa_{pq}^{*}} = \sum_{rs} \kappa_{rs} \ERI{pq}{rs} 
    \label{eq:Delta}
\end{equation}
where the Hartree and exchange potentials are
\begin{subequations}
\begin{align}
	J_{pq} & =  \sum_{rs} D_{rs} \ERI{pr}{qs}
	\\
	K_{pq} & =  \sum_{rs} D_{rs} \ERI{pr}{sq}
\end{align}
\end{subequations}
This allows us to define the HFB generalized Fock matrix (for spin-unpolarized systems) as
\begin{equation} \label{eq:H_su}
	\bH = 
	\begin{pmatrix}
		\bF & \bDelta \\
		\bDelta^* & -\bF^*
	\end{pmatrix}
\end{equation}

This Hamiltonian is Hermitian and can therefore be diagonalized to yield real eigenvalues that occur in pairs, $\pm \beps$, as a direct consequence of particle-hole symmetry.
The paired structure is immediately apparent in the special case $\bDelta = \bO$, where the Hamiltonian reduces to a block-diagonal form with $\pm \bF$ on the diagonal. 
In this limit, $\bF$ coincides with the usual Hartree--Fock operator, and the $\pm$ symmetry of the spectrum follows directly.
The eigenvalue problem can thus be formulated as
\begin{equation}
    \bH
    \cdot
    \begin{pmatrix}
		\bV^* & \bU \\
		\bU^* & \bV
    \end{pmatrix}
	=
    \begin{pmatrix}
		\bV^* & \bU \\
		\bU^* & \bV
    \end{pmatrix}
    \cdot
    \begin{pmatrix}
		-\beps & \bO \\
		\bO & \beps
    \end{pmatrix}
\end{equation}
where the matrices $\bU$ and $\bV$ are constructed from the eigenvectors of the Hamiltonian~\footnote{Note that the eigenvectors associated with positive quasiparticle energies are related to those associated with negative energies through the relation
$ \begin{pmatrix}
		 \bU \\
		 \bV
    \end{pmatrix}
= \begin{pmatrix}
		{\bf 0} & {\bf -1} \\
		{\bf 1} & {\bf 0}
    \end{pmatrix}
    \cdot
 \begin{pmatrix}
		\bV^*  \\
		\bU^* 
    \end{pmatrix}^* $ }. In spin-unpolarized systems they adopt the following block spin structure:
\begin{align}
    \bV & = 
    \begin{pmatrix}
		\bV^\upar & \bO \\
		\bO & \bV^\dwar
	\end{pmatrix}
	&
    \bU & = 
	\begin{pmatrix}
		\bO & \bU^\upar \\
		\bU^\dwar & \bO
	\end{pmatrix},
\end{align}
where we reincorporated the spin structure of the full matrix for clarity. In addition, the normal and anomalous density matrices can be expressed as~\cite{ring2004nuclear}
\begin{align}
	\bD & = \bV^* \cdot \T{\bV}
	&
	\bkappa & = \bV^* \cdot \T{\bU}
\end{align}
where the matrices $\bU$ and $\bV$ satisfy the Bogoliubov orthonormality conditions
\begin{align}
	{\bU}^\dagger \cdot \bU + {\bV}^\dagger \cdot \bV & = \bI
	&
	\T{\bU} \cdot \bV + \T{\bV} \cdot \bU & = \bO
\end{align}
which ensures preservation of the fermionic anticommutation relations~\cite{ring2004nuclear,staroverov2002optimization}. 
The matrices $\bU$ and $\bV$ enter the definition of the unitary matrix $\bW$ [see Eq.~\ref{eq:bogoliubov_trans}], which defines the Bogoliubov transformation required to express the operators in the quasiparticle representation. 
Applying this transformation to the generalized density matrix $\bR$, one obtains
\begin{equation} \label{eq:R_qp_basis}
    \bW^\dagger \cdot \bR \cdot \bW =
	\begin{pmatrix}
		\bO & \bO\\
		\bO & \bI \\
	\end{pmatrix},
\end{equation}
which corresponds to setting $D_{PQ}=\mel{\Phi_{\HFB}}{\cre{\alpha}{Q}\anh{\alpha}{P}}{\Phi_{\HFB}}$ and $\kappa_{PQ} = \mel{ \Phi_{\HFB} }{ \anh{\alpha}{Q} \anh{\alpha}{P} }{  \Phi_{\HFB} }$~\footnote{Recalling that $\anh{\alpha}{P} \ket{\Phi_{\HFB} } =0$ for all $P$, it is straightforward to conclude that the elements $D_{PQ}=\mel{\Phi_{\HFB}}{\cre{\alpha}{Q}\anh{\alpha}{P}}{\Phi_{\HFB}}$ and $\kappa_{PQ} = \mel{ \Phi_{\HFB} }{ \anh{\alpha}{Q} \anh{\alpha}{P} }{  \Phi_{\HFB} }$ are all zero in the quasiparticle basis. Thus, only the identity matrix in the definition of $\bR$ remains [see Eq.~(7.29) of Ref.~\cite{ring2004nuclear} for more details].}. When $\bR$ is expressed in this form, it is said to be in the canonical representation~\cite{ring2004nuclear}. 
In this representation, the idempotency of $\bR$ becomes immediately apparent.

Note that all the states with negative quasiparticle energies are employed in the construction of the density matrices. 
This corresponds to defining the vacuum of the quasiparticles [see Eq.~\eqref{eq:vacuum_qp}] that minimizes the expectation value of the Hamiltonian and thus defines the HFB ground state. 
In other words, the ground state is obtained by occupying all negative-energy quasiparticle modes, in direct analogy with the Aufbau principle in HF theory. 
With this convention, the generalized density matrix becomes a projector onto the occupied quasiparticle space, which guarantees its idempotency and leads to the canonical block structure shown in Eq.~\eqref{eq:R_qp_basis}. 

Because the Hamiltonian depends explicitly on both the density and anomalous density matrices, the procedure must be carried out iteratively. 
The self-consistent HFB algorithm is described in detail in Appendix \ref{app:hfb_algo}.
Convergence is typically assessed through the commutator condition $\comm{ \bH }{ \bR} = \bO$. 

Interestingly, for purely repulsive electronic interactions, it can be shown that the pairing contribution [the last term in Eq.~\eqref{eq:E_bcs_su}] is strictly positive and therefore increases the total energy~\cite{bardeen1957theory,staroverov2002optimization}. 
As a result, the variational minimum is obtained for a vanishing anomalous density ($\bkappa=\bO$).
In this limit, Eq.~\eqref{eq:an_den_rel} reduces to $\bD^2 = \bD$, which implies that the density matrix is idempotent. 
The generalized density matrix thus becomes a projector onto an occupied one-particle subspace, and the HFB wavefunction collapses to a single Slater determinant. 
Consequently, the HFB equations usually reduce to the standard HF equations (with $\mu$ defining the Fermi level)~\cite{duguet2020zero}. 
This clearly shows that an attractive interaction is essential for superconductivity: only when the pairing channel lowers the total energy can a nonzero anomalous density develop, leading to a genuine HFB state that differs from a single-determinant HF solution.

\subsection{The Bogoliubov $GW$ approximation}
\label{subsec:bogoliubov_gw}
The $GW$ approximation is typically introduced by first defining the exact one-body Green's function (or propagator)~\cite{martin2016interacting}
\begin{equation} \label{eq:G_he}
    G^\he(12)=-\ii \mel{ \Psi_0^N }{ \mathcal{T}[\anh{\psi}{}(1)\cre{\psi}{}(2)] }{ \Psi_0^N }
\end{equation}
where the composite index $1=(\bx_1,t_1)=(\br_1,\sigma_1,t_1)$ accounts for the position, spin, and time coordinates, $\Psi_0^N$ is the exact ground-state many-body wavefunction of the $N$-electron system, $\mathcal{T}$ is the time-ordering operator, and $\cre{\psi}{} (\anh{\psi}{})$ is a time-dependent creation (annihilation) field operator. 

Physically, the Green's function represents the propagation of electrons and holes in the system~\cite{martin2016interacting}. 
In the spin-orbital basis $\{\phi_P\}$, it can be expanded as
\begin{equation}
  G^\he(12) = \sum_{PQ} \phi_P(\bx_1) G^{\he}_{PQ}(\tau) \phi_Q^*(\bx_2)
\end{equation}
with $\tau =t_1-t_2$ and
\begin{equation}
\begin{split}
	G^{\he}_{PQ}(\tau) 
	& =-\ii \theta(\tau) \sum_m e^{-\ii (E^{N+1}_m-E_0^N)\tau} \mel*{\Psi_0^N}{\anh{a}{P}}{\Psi^{N+1}_m}\mel*{\Psi^{N+1}_m}{\cre{a}{Q}}{\Psi_0^N}  
	\\ 
 	& +\ii \theta(-\tau) \sum_m e^{-\ii (E_0^N-E^{N-1}_m)\tau} \mel*{\Psi_0^N}{\cre{a}{Q}}{\Psi^{N-1}_m}\mel*{\Psi^{N-1}_m}{\anh{a}{P}}{\Psi_0^N}
\end{split}
\end{equation}
where $\theta(x)$ is the Heaviside step function, $E_0^N = \mel*{\Psi_0^N}{\hH}{\Psi_0^N}$ is the ground-state energy of the $N$-electron system, and $\Psi^{N\pm1}_m$ and $E^{N\pm1}_m$ denote the eigenstates and eigenenergies of the $(N\pm1)$-electron systems, respectively. 
The first term describes electron propagation (addition processes), while the second corresponds to hole propagation (removal processes).

An important property follows directly from this expression. 
Taking the limit $\tau \to 0^-$, the Green's function yields the one-body density matrix~\cite{martin2016interacting}
\begin{equation}
	D_{PQ}= - \ii G^{\he}_{PQ} (\tau \to 0^-)
\end{equation}
This property is widely exploited in the literature to compute correlated density matrices and to develop new methods~\cite{bruneval2019improved,bruneval2019assessment,bruneval2021improved,denawi2023gw}. 

The Green's function defined in Eq.~\eqref{eq:G_he} can be evaluated using $\Phi_{\HFB}$, and the one-body density matrix can, in principle, be extracted from it.
However, to capture all one-body correlations contained in $\Phi_{\HFB}$, it is necessary to introduce the hole-hole (hh), electron-electron (ee), and electron-hole (eh) propagators:
\begin{subequations}
\begin{align}
	G^\hh(12) & =-\ii \mel{\Psi_0}{\mathcal{T}[\anh{\psi}{}(1)\anh{\psi}{}(2)]}{\Psi_0}
	\label{eq:G_hh}  
	\\
	G^\ee(12) & =-\ii \mel{\Psi_0}{\mathcal{T}[\cre{\psi}{}(1)\cre{\psi}{}(2)]}{\Psi_0}
	\label{eq:G_ee}  
	\\
	G^\eh(12) & =-\ii \mel{\Psi_0}{\mathcal{T}[\cre{\psi}{}(1)\anh{\psi}{}(2)]}{\Psi_0}
	\label{eq:G_eh}  
\end{align}
\end{subequations}
with $G^{\eh}(12)=-G^{\he}(21)$.
Because $G^\ee$ and $G^\hh$ do not conserve the particle number, they are usually referred to as anomalous (or pairing) propagators.

Using Nambu's formalism~\cite{nambu1960quasi}, these propagators can be combined into the Gorkov Green's function~\cite{gor1958energy}:
\begin{equation} \label{eq:G_gorkov}
	\bG(12) = 
	\begin{pmatrix}
		G^\he(12) &  G^\hh(12) \\
		G^\ee(12) & G^\eh(12)
	\end{pmatrix}
\end{equation}
Expanding the field operators in the spin-orbital basis and introducing the corresponding matrix elements $G_{PQ}(\tau)$, one finds that in the limit $\tau \to 0^-$ the Gorkov Green's function reduces to the generalized density matrix $\bR$
\begin{equation}
	R_{PQ}= - \ii G_{PQ} (\tau \to 0^-)
\end{equation}
thereby providing direct access to both the normal density matrix $\bD$ and the pairing tensor $\bkappa$, i.e., to all one-body information encoded in the HFB reference state.

It can be shown that, $\bG$ has the same block structure as $\bR$ in Eq.~\eqref{eq:R_full_so_false}. 
In particular, the elements of the Gorkov propagator that lead to the outer density matrices of $\bR$ in Eq.~\eqref{eq:R_full_so_false}, (i.e., $\bD^{\upar\upar}$, $\bkappa^{\upar\dwar}$, $\bkappa^{\dwar\upar}$, and $\bD^{\dwar\dwar}$) can be written as
\begin{subequations}
\begin{align}
		[\bG_0^{\he}(\tau)]^{\upar\upar}
		= - \ii \theta(\tau) \bU^\upar \cdot e^{-\ii\beps\tau} \cdot (\bU^\upar)^\dag  
		+ \ii \theta(-\tau) (\bV^\upar)^* \cdot e^{\ii\beps\tau} \cdot (\bV^\upar)^T
	\label{eq:G_he_t}
	\\
		[\bG_0^{\hh}(\tau)]^{\upar\dwar}
		= - \ii \theta(\tau) \bU^\upar \cdot e^{-\ii\beps\tau} \cdot (\bV^\dwar)^\dag 
		+ \ii \theta(-\tau) (\bV^\upar)^* \cdot e^{\ii\beps\tau} \cdot (\bU^\dwar)^T
	\label{eq:G_hh_t}
	\\
		[\bG_0^{\ee}(\tau)]^{\dwar\upar}
		= - \ii \theta(\tau) \bV^\dwar \cdot e^{-\ii\beps \tau} \cdot (\bU^\upar)^\dag
		+ \ii \theta(-\tau) (\bU^\dwar)^* \cdot e^{\ii\beps\tau} \cdot (\bV^\upar)^T 
	\label{eq:G_ee_t}
	\\
		[\bG_0^{\eh}(\tau)]^{\dwar\dwar}
		= - \ii \theta(\tau) \bV^\dwar \cdot e^{-\ii\beps\tau} \cdot (\bV^\dwar)^\dag 
		+ \ii \theta(-\tau) (\bU^\dwar)^* \cdot e^{\ii\beps\tau} \cdot (\bU^\dwar)^T 
	\label{eq:G_eh_t}
\end{align}
\end{subequations}
Here, the subscript $0$ indicates that the propagator is evaluated from the HFB reference wavefunction, and $e^{\pm \ii\beps\tau}$ denotes a diagonal matrix built using the HFB quasiparticle energies.
For spin-unpolarized calculations, the missing propagators can be obtained by exchanging spin-up and spin-down indices in Eqs.~\eqref{eq:G_he_t}--\eqref{eq:G_eh_t}.

Let us emphasize that, for spin-unpolarized systems, only the Gorkov propagator components introduced in Eqs.~\eqref{eq:G_he_t}--\eqref{eq:G_eh_t} are required. 
In this case, one can define the spin-unpolarized Gorkov propagator as
\begin{equation} \label{eq:G_gorkov_su}  
	\bG_0(\tau)=
	\begin{pmatrix}
		\bG_0^{\he}(\tau) &  \bG_0^{\hh}(\tau) \\
		\bG_0^{\ee}(\tau) &  \bG_0^{\eh}(\tau)
	\end{pmatrix}
\end{equation}
where each block implicitly refers to a single spin channel. 
This restriction significantly simplifies the numerical implementation without loss of generality for spin-unpolarized systems.

Finally, for practical calculations (see Sec.~\ref{sec:scgw_algo}), it is convenient to evaluate the propagators in imaginary time ($\ii\tau$)~\footnote{In practice, the usual convention is to evaluate $-\ii \tau$ in Eqs.~\eqref{eq:G_he_t}--\eqref{eq:G_eh_t}.}. In that case, for example, the hole-electron elements are given by
\begin{equation} \label{eq:G_he_t_im}
	\bG_0^{\he}(\ii \tau)
	= - \ii \theta(\tau) \bU \cdot e^{-\beps\tau} \cdot (\bU)^\dag  
	+ \ii \theta(-\tau) (\bV)^* \cdot e^{\beps\tau} \cdot (\bV)^T
\end{equation}
Similarly, the $\hh$, $\ee$, and $\eh$ propagators can be evaluated in imaginary time, yielding the same type of exponential structure in their matrix elements.

It is also useful to work in the imaginary-frequency ($\ii\omega$) domain via the Fourier transform:
\begin{equation}
	\bG(\ii\omega) = -\ii \int \dd{\tau} \bG(\ii\tau) e^{-\ii\omega \tau}
\end{equation}
with the inverse transform
\begin{equation}
	\bG(\ii\tau) = \frac{\ii}{2\pi} \int \dd{\omega} \bG(\ii\omega) e^{\ii\omega \tau}
\end{equation}
These relations allow the Gorkov propagator to be explicitly transformed between the imaginary-time and imaginary-frequency representations. Note that in the frequency domain, the density matrix can be computed as discussed in Appendix~\ref{app:grids}.

The Gorkov propagator corresponding to the HFB wavefunction describes an uncorrelated system. 
Correlation effects can be included via the Dyson equation, which relates the fully correlated propagator to the uncorrelated one:
\begin{align} \label{eq:Dyson}
	\bG(11') = \bG_0(11') + \int \dd{2}\dd{2'} \bG_0(12) \cdot \bSig(22') \cdot \bG(2'1')
\end{align}
where the Gorkov self-energy~\cite{essenberger2014density}
\begin{equation} \label{eq:Sigma_gorkov}
    \bSig(12) =
    	\begin{pmatrix}
    	\bSig^{\he}(12) &  \bSig^{\hh}(12) \\
    	\bSig^{\ee}(12) &  \bSig^{\eh}(12)
    \end{pmatrix}
\end{equation}
depends explicitly on the two-body Green's function. 
In practice, this dependence is commonly simplified using the Schwinger relation~\cite{martin1959theory}, which allows the Gorkov self-energy to be expressed as a functional of the one-body Green's function. 
For further details, see Refs.~\cite{essenberger2014density,linscheid2015hedin,marie2024anomalous,marie2025anomalous,lane2026gorkov}.

Using many-body perturbation theory, the Gorkov self-energy can, in principle, be expressed exactly. 
In practice, however, $\bSig(12)$ must be approximated~\cite{martin2016interacting} by truncating the perturbative expansion, as illustrated in Refs.~\cite{soma2020self,soma2011ab}. 
In this work, we adopt the $GW$ approximation~\cite{hedin1965new} for the Gorkov self-energy: 
\begin{equation} \label{eq:Sigma_gorkov_gw}
	\bSig(11')=\ii
	\int \dd{3} \dd{3'} 
	\begin{pmatrix}
		G^{\he}(33') &   -G^{\hh}(33') \\
		-G^{\ee}(33') &  G^{\eh}(33')
	\end{pmatrix}
	W(11';33')
\end{equation}    
where $W$ is the so-called screened Coulomb interaction~\cite{martin2016interacting}. 

To define the screened Coulomb interaction, let us first introduce the non-interacting polarizability, which, within the $GW$ approximation~\cite{essenberger2014density,linscheid2015hedin,marie2024anomalous,marie2025anomalous,lane2026gorkov}, reads
\begin{equation}
    \chi_0 (12) = -\ii \qty[ G^{\he}(21) G^{\he}(12)-G^{\hh}(21) G^{\ee}(12) ]
\end{equation}
Then, making use of the symmetry properties of the $\hh$ and $\ee$ Green's functions, we can express $\chi_0$ in a basis representation, which leads to the matrix elements
\begin{subequations}
\begin{align}
   [ \bchi _0]^{\upar\upar,\upar\upar}_{rq,sp} (\ii\tau ) & = -\ii [\bG^{\he}(\ii\tau)]^{\upar\upar}_{pq} [\bG^{\he}(-\ii\tau)]^{\upar\upar}_{rs}
   \label{eq:xo_ubar}
   \\
   [ \bchi _0]^{\dwar\dwar,\dwar\dwar}_{rq,sp} (\ii\tau ) & = -\ii [\bG^{\he}(\ii\tau)]^{\dwar\dwar}_{pq} [\bG^{\he}(-\ii\tau)]^{\dwar\dwar}_{rs}
   \label{eq:xo_bar}
   \\
   [ \bchi _0]^{\dwar\dwar,\upar\upar}_{rq,sp} (\ii\tau ) & = \ii [\bG^{\hh}(\ii\tau)]^{\upar\dwar}_{pq} [\bG^{\ee}(-\ii\tau)]^{\dwar\upar}_{rs}
   \label{eq:xo_hhee_1}
   \\
   [ \bchi _0]^{\upar\upar,\dwar\dwar}_{rq,sp} (\ii\tau ) & = \ii [\bG^{\hh}(\ii\tau)]^{\dwar\upar}_{pq} [\bG^{\ee}(-\ii\tau)]^{\upar\dwar}_{rs}
   \label{eq:xo_hhee_2}
\end{align}
\end{subequations}
The indices can be combined to construct a conventional matrix representation.
In general, the polarizability is an even function of time, so it is sufficient to evaluate it for $\tau > 0$. 
For spin-unpolarized systems, it is straightforward to show that Eqs.~\eqref{eq:xo_ubar} and \eqref{eq:xo_bar}, as well as Eqs.~\eqref{eq:xo_hhee_1} and \eqref{eq:xo_hhee_2}, are equivalent.

Analyzing the contributions to the polarizability, one sees that only the matrix elements in Eqs.~\eqref{eq:xo_ubar} and \eqref{eq:xo_bar} remain non-zero when the HFB method reduces to HF (i.e., when the anomalous density vanishes). 
These correspond to ``same-spin'' contributions. 
In contrast, the anomalous density contributions in Eqs.~\eqref{eq:xo_hhee_1} and \eqref{eq:xo_hhee_2} correspond to ``opposite-spin'' contributions, analogous to those encountered in relativistic two-component calculations expressed in the atomic orbital basis~\cite{aryasetiawan2008generalized,forster2023two,abraham2024relativistic}. 
For further details, see Ref.~\cite{lane2026gorkov}. 

Once the non-interacting polarizability $\bchi_0(\ii\tau)$ is computed, it can be Fourier-transformed to the imaginary-frequency domain, $\bchi_0(\ii\omega)$, to construct the interacting polarizability within the random-phase approximation (RPA)~\cite{martin2016interacting,co2023introducing} 
\begin{equation}
    \bchi (\ii\omega) = \qty[ \bI-\bchi_0(\ii\omega) \cdot \bv ]^{-1} \cdot \bchi_0(\ii\omega)
    \label{eq:Polarizability}
\end{equation}
where the matrix elements of the bare Coulomb interaction are $v_{RP,SQ}= \braket{PQ}{RS}$. Note that the spin-structure of $\bchi$ is the same as that of $\bchi_0$.

Within the $GW$ approximation, the screened Coulomb interaction is defined from the polarizability as~\cite{martin1959theory,hedin1965new,marie2024anomalous,marie2025anomalous,essenberger2014density}
\begin{equation}
\begin{split}
    \bW(\omega) 
    &= \bv + \bv \cdot \bchi_0(\omega) \cdot \bW(\omega) \\
    &= \bv + \bv \cdot \bchi(\omega) \cdot \bv ,
\end{split}
\end{equation}
where $\bchi$ is the interacting polarizability. 
The screened interaction is commonly decomposed into a static contribution $\bv$ and a dynamic component
\begin{equation}
    \bW_p(\ii\omega) = \bv \cdot \bchi(\ii\omega) \cdot \bv
\end{equation}
The static term, when inserted into Eq.~\eqref{eq:Sigma_gorkov_gw}, reproduces the exchange contribution of the Fock operator [see Eq.~\eqref{eq:Fock}] as well as the pairing potential [see Eq.~\eqref{eq:Delta}], which are already included at the HFB level (see Refs.~\cite{essenberger2014density,marie2024anomalous,marie2025anomalous} for details). 
The dynamic part, on the other hand, accounts for correlation effects. 
It can be inverse Fourier transformed to the imaginary-time domain, yielding $\bW_p(\ii\tau)$. 
Since the screened interaction inherits the symmetry properties of the polarizability, it is an even function of both $\tau$ and $\omega$.

It is worth noting that, in textbooks, $W$ is often introduced through a nonlocal, frequency-dependent dielectric function, $\epsilon^{-1}$~\cite{martin2016interacting}, which accounts for electron screening. 
This screening effectively modifies the electron-electron interaction, altering the two-electron integrals accordingly.

Finally, $\bW_p(\ii\tau)$ can be inserted in Eq.~\eqref{eq:Sigma_gorkov_gw} to produce the correlation part of the self-energy $\bSig_\text{c}$, whose matrix elements are given as
\begin{subequations}
\begin{align}
    [\bSig_\text{c}^\he (\ii\tau)]_{pq}^{\upar\upar} & = \ii \sum_{rs} [\bW_p (\ii\tau)]_{rp,qs}^{\upar\upar,\upar\upar} [\bG^{\he} (\ii\tau)]_{rs}^{\upar\upar}
    \label{eq:Sigma_he_c}
\\
    [\bSig_\text{c}^\hh (\ii\tau)]_{pq}^{\upar\dwar} & = -\ii \sum_{rs} [\bW_p (\ii\tau)]_{rp,sq}^{\upar\upar,\dwar\dwar} [\bG^{\hh} (\ii\tau)]_{rs}^{\upar\dwar}
    \label{eq:Sigma_hh_c}
\\
    [\bSig_\text{c}^\ee (\ii\tau)]_{pq}^{\dwar\upar} & =- \ii \sum_{rs} [\bW_p (\ii\tau)]_{pr,qs}^{\dwar\dwar,\upar\upar} [\bG^{\ee} (\ii\tau)]_{rs}^{\dwar\upar}
    \label{eq:Sigma_ee_c}
\\
    [\bSig_\text{c}^\eh (\ii\tau)]_{pq}^{\dwar\dwar} & = \ii \sum_{rs} [\bW_p (\ii\tau)]_{pr,sq}^{\dwar\dwar,\dwar\dwar}  [\bG^{\eh} (\ii\tau)]_{rs}^{\dwar\dwar}
    \label{eq:Sigma_eh_c}
\end{align}
\end{subequations}
Again, for spin-unpolarized calculations, the missing terms are equivalent to their corresponding spin-up or spin-down counterparts, i.e., $[\bSig_\text{c}^\he (\ii\tau)]^{\dwar\dwar}=[\bSig_\text{c}^\he (\ii\tau)]^{\upar\upar}$, $[\bSig_\text{c}^\hh (\ii\tau)]^{\dwar\upar}=-[\bSig_\text{c}^\hh (\ii\tau)]^{\upar\dwar}$, etc. 

In general, the self-energy is not an even function of $\tau$; therefore, it must also be constructed explicitly for $\tau<0$. 
Since the screened Coulomb interaction is even in $\tau$, it is sufficient to evaluate only the propagators for $\tau<0$ in Eqs.~\eqref{eq:Sigma_he_c}--\eqref{eq:Sigma_eh_c} to obtain $\bSig_\text{c}(\ii\tau)$ in this regime. 
Once built, $\bSig_\text{c}(\ii\tau)$ can be inserted into Eq.~\eqref{eq:Dyson} to obtain the correlated Gorkov propagator.

In practice, it is more convenient to first Fourier transform the self-energy to the imaginary-frequency domain and compute the correlated propagator as
\begin{equation} \label{eq:Dyson_w}
	\bG(\ii\omega) = \qty[ \ii\omega \bI - \bH - \bSig_\text{c}(\ii\omega) ]^{-1}
\end{equation}
which is simply the imaginary-frequency form of the Dyson equation, with the uncorrelated propagator given by
\begin{equation} \label{eq:Dyson_w_g0}
	\bG_0(\ii\omega) = \qty[ \ii\omega \bI - \bH ]^{-1}
\end{equation}

For our spin-unpolarized calculations, only the matrix elements of $\bSig_\text{c}$ defined in Eqs.~\eqref{eq:Sigma_he_c}–\eqref{eq:Sigma_eh_c}, together with $\bH$ as the Hamiltonian and $\bG$ as the Gorkov propagator, are required. These will be used directly in Eqs.~\eqref{eq:Dyson_w} and~\eqref{eq:Dyson_w_g0} in our numerical implementations. 

Constructing a correlated Gorkov propagator provides access to a correlated generalized density matrix $\bR$, and thus to both $\bD$ and $\bkappa$. 
However, since $\bH$ depends explicitly on the normal and anomalous densities and $\bSig_\text{c}$ depends on the Gorkov propagator itself, obtaining the correlated $\bG$ requires a self-consistent procedure~\cite{hedin1965new}. 
This iterative process defines the Bog.~sc$GW$ method evaluated in this work.

In practice, the procedure involves Fourier-transforming the correlated $\bG(\ii\omega)$ back to the imaginary-time domain to obtain $\bG(\ii\tau)$, which is then used in Eqs.~\eqref{eq:xo_ubar}--\eqref{eq:xo_hhee_2} to update $\bchi_0$ and other quantities within the iteration. 
During this process, the chemical potential must be adjusted as needed to ensure that the final Gorkov propagator reproduces the target average particle number.

Once convergence is reached, the correlated Gorkov propagator can also be used to evaluate the total electronic energy via, for example, the Galitskii--Migdal (GM) functional~\cite{galitskii1958application}:
\begin{equation}
    E_{\GM}[\bG] = E_\HFB[\bD,\bkappa] + E_\text{c}^{\GM}[\bG]
\end{equation}
where $E_\HFB[\bD,\bkappa]$ is evaluated with the converged densities, and the Galitskii-Migdal correlation energy functional reads
\begin{equation}
    E_\text{c}^{\GM}[\bG] = \frac{1}{4\pi} \int_{-\infty}^{\infty} \dd{\omega}
   \Tr \qty{ \bv \cdot \qty[ \bchi_0(\ii \omega) - \bchi(\ii \omega) ] }
   \label{eq:GMc}
\end{equation}
with $\Tr$ denoting the trace of the matrix. 

\rev{Since sc$GW$ is a $\Phi$-derivable approximation, at full self-consistency the GM, Klein, and Luttinger--Ward functionals yield the same total energy~\cite{caruso2012unified,caruso2013self}; we therefore use the GM functional without loss of generality.}

\rev{For Bog.~scGF2, the same algorithm is used with $\bSig_\text{c}$ replaced by its second-order expression (see Section~II of the SI), and the total energy is again obtained as $E=E_\HFB[\bD,\bkappa]+E_\text{c}^{\GM}$, with $E_\text{c}^{\GM}[\bG]=-\int\dd{\tau}\,\Tr\qty{\qty[\bSig_\text{c}(-\ii\tau)\,\bG(\ii\tau)]^{\he}}$, where $[\cdots]^{\he}$ denotes the upper-left ($\he$) block of the Gorkov matrix product, and the prefactor results from the usual factor $1/2$ of the GM functional times a factor 2 accounting for the two spin channels. The complete algorithm is given in Section~IV of the SI.}

Finally, when the single-particle states are real-valued, the eigenvectors of $\bH$ are also real, as are $\bR$, $\bD$, and $\bkappa$. 
This simplification allows several $GW$ quantities to be expressed more efficiently. 
For example, the matrix elements of the non-interacting polarizability reduce to
\begin{multline}\label{eq:xo_real}
   [ \bchi_0]^{r q}_{s p} (\ii\tau ) 
   = -2\ii \left( [\bG^{\he}(\ii\tau)]_{p q} [\bG^{\he}(-\ii\tau)]_{r s} \right. 
   \\
   \left. -  [\bG^{\hh}(\ii\tau)]_{p q} [\bG^{\ee}(-\ii\tau)]_{r s} \right)
\end{multline}   
where the factor of 2 accounts for the two spin channels. 
Importantly, this expression can be evaluated using only a single spin channel, which reduces the computational cost.

\section{The Bogoliubov $\text{sc}GW$ algorithm}
\label{sec:scgw_algo}

In this section, we briefly describe the self-consistent procedure implemented in this work. 
All algorithms have been incorporated into the \textsc{quack} package (\url{https://github.com/pfloos/QuAcK.git}), our in-house electronic-structure code.

Once the HFB self-consistent procedure is completed (see Appendix~\ref{app:hfb_algo}), the quasiparticle amplitudes $\bU$, $\bV$, and energies $\beps$ are used to construct the HFB generalized density matrix $\bR$ and the non-interacting Gorkov propagator $\bG_0$. 
The latter serves as the starting point of the Bog.~sc$GW$ procedure. 
In practice, we construct $\bG_0(\pm \ii\tau_\kappa)$ on an imaginary-time grid with roots $\tau_\kappa$ and, for convenience, also evaluate $\bG_0(\ii\omega_\iota)$ on a corresponding imaginary-frequency grid with roots $\omega_\iota$. 
The frequency representation facilitates the evaluation of $\bR$ from the propagator, as discussed below.
We refer the interested reader to Appendix~\ref{app:grids} for additional details about the quadrature grids employed in the present work.

Expressions for $\bG_0(\ii\omega)$ are provided in Appendix~\ref{app:gorkov_freq}. 
Alternatively, Eq.~\eqref{eq:Dyson_w_g0} can be used directly. 
From $\bG_0$, we construct $\bchi_0$, $\bW_p$, and $\bSig_\text{c}$, as described in Sec.~\ref{subsec:bogoliubov_gw}, and then obtain the correlated Gorkov propagator $\bG(\ii\omega)$ via the Dyson equation [see Eq.~\eqref{eq:Dyson_w}]. 
The correlated propagator is subsequently used to build the generalized density matrix $\bR$.

However, the slow asymptotic decay of $\bG(\ii\omega)$ ($\sim \omega^{-1}$) complicates the numerical quadrature required to compute $\bR$ [see Eq.~\eqref{eq:R_from_G}]. 
To alleviate this issue, and following Refs.~\cite{shishkin2007accurate,grumet2018beyond}, we introduce the difference
\begin{equation}
    \Delta \bG(\ii\omega) = \bG(\ii\omega) -\bG_0(\ii\omega)
\end{equation}
and compute instead
\begin{equation}
    \Delta\bR \approx \frac{1}{\pi}\sum ^{N_\omega} _\kappa \eta_\kappa \text{Re} \qty[ \Delta\bG (\ii\omega_\kappa) ]
        \label{eq:DR_from_DG}
\end{equation}
so that the correlated generalized density matrix is obtained as
\begin{equation} \label{eq:R_fromG}
    \bR = \bR_\HFB + \Delta \bR
\end{equation}
This strategy is advantageous because $\Delta \bG(\ii\omega)$ decays as $(\ii\omega)^{-3}$, which significantly improves the convergence of the numerical integration along the imaginary axis.

From the updated $\bR$, we extract the normal density matrix $\bD$ and verify its trace to determine whether the chemical potential $\mu$ must be adjusted. 
If necessary, $\mu$ is refined (for example, using the bisection method~\cite{burden1997numerical}) while keeping $\bD$, $\bkappa$, and $\bSig_\text{c}(\ii\omega)$ fixed, since only the Hamiltonian $\bH$ depends explicitly on $\mu$. 
Equation~\eqref{eq:Dyson_w} is then used to recompute $\bG(\ii\omega)$ and the corresponding densities.

Once the trace condition for $\bD$ is satisfied, the updated $\bD$ and $\bkappa$ (extracted from $\bR$) are employed to recompute the Fock matrix $\bF$ and the pairing matrix $\bDelta$ according to Eqs.~\eqref{eq:Fock} and \eqref{eq:Delta}. 
These matrices define a new Hamiltonian $\bH$, which is subsequently used to construct an updated $\bG(\ii\omega)$ [via Eq.~\eqref{eq:Dyson_w}] and a new $\bR$ (with further adjustment of $\mu$ if required).

This self-consistent cycle is repeated until the change in $\bG(\ii\omega)$ and/or $\bR$ between successive iterations falls below a predefined convergence threshold $\varpi_1$. 
Only once $\bG(\ii\omega)$ and/or $\bR$ have converged can the inverse Fourier transform be applied to obtain the imaginary-time propagator $\bG(\ii\tau)$.
However, this transformation again presents numerical challenges due to the slow asymptotic decay of $\bG(\ii\omega)$. 
For this reason, it is preferable to perform the transformation on the difference
\begin{equation}
    \Delta\bG (\pm\ii\tau_\kappa) \approx  \ii \sum ^{N_\omega} _\iota \left[ \eta^{\cos}_{\iota,\kappa} \Re \left[ \Delta\bG (\ii\omega_\iota)\right] \mp \eta^{\sin}_{\iota,\kappa} \Im \left[ \Delta\bG (\ii\omega_\iota)\right]\right] 
        \label{eq:DGtP_from_DGw}
\end{equation}
where $\eta^{\cos}_{\iota,\kappa}$ and $\eta^{\sin}_{\iota,\kappa}$ denote the cosine and sine quadrature weights generated by the \texttt{greenX}~\cite{azizi2024validation,azizi2023time} library (see Appendix~\ref{app:grids}).

The correlated Gorkov propagator in imaginary time is then reconstructed as
\begin{equation} \label{eq:Ggorkov_from_itau_1}
	\bG (\pm\ii\tau_\kappa)=\bG_0 (\pm\ii\tau_\kappa)+\Delta\bG (\pm\ii\tau_\kappa)   
\end{equation}
By working with $\Delta\bG$ rather than $\bG$ directly, the improved high-frequency decay ($\sim \omega^{-3}$) ensures better numerical stability and accuracy in the transformation between the imaginary-frequency and imaginary-time domains.

The correlated propagator is subsequently employed to construct updated quantities $\bchi_0$, $\bW_p$, and $\bSig_\text{c}(\ii\omega)$ for the next macro-iteration.
The self-consistent cycle is terminated when the change in $\bG(\ii\omega)$ and/or $\bR$ between two consecutive macro-iterations falls below a predefined threshold, $\varpi_2$.
The complete Bog.~sc$GW$ procedure is summarized in Algorithm~\ref{alg:algo_bscgw}. 
To accelerate convergence and enhance numerical stability, a direct inversion in the iterative subspace (DIIS) scheme~\cite{pulay1980convergence} is applied to both $\bR$ and $\bG(\ii\tau)$ throughout the self-consistent cycle. It is worth mentioning that the Fourier transforms involving $\bchi_0$, $\bW_p$, and $\bSig_\text{c}(\ii\omega)$ do not suffer from slow decays of the integrands and can be efficiently performed with grids and weights imployed in this work.

Finally, regarding the computational cost of the Bog.~sc$GW$ method, the principal bottleneck arises from the matrix inversion required to construct $\bchi(\ii\tau)$, which leads to a formal scaling of $\order*{M^6}$ with respect to the system size $M$. 
Likewise, the construction of $\bW(\ii\omega)$, obtained as the product of $\bchi(\ii\omega)$ and the bare interaction matrix $\bv$, exhibits the same formal scaling.
It is well established, however, that the use of density fitting (or resolution-of-the-identity) techniques can substantially reduce the cost of these operations to an effective cubic scaling with system size~\cite{whitten1973coulombic,duchemin2021cubic,wilhelm2021low,shishkin2007self,forster2021low,yeh2022fully}. 
This reduction represents a key practical advantage and significantly enhances the applicability of the method to larger systems.
In the present work, our in-house implementation does not employ density fitting. 
Consequently, the reported calculations follow the formal $\order*{M^6}$ scaling of the underlying algorithm.

\begin{algorithm}[H]
\begin{algorithmic}
	\Procedure{Bogoliubov sc$GW$}{$\bU$,$\bV$,$\beps$,$\mu$}
		\State Get $\eta_\kappa,\eta^{\cos}_{\kappa,\iota},\eta^{\sin}_{\kappa,\iota},\eta^{\cos}_{\iota,\kappa},\eta^{\sin}_{\iota,\kappa},\tau_k,\omega_k$ from \texttt{greenX}  
		\State Construct $\bG_0(\ii\omega_\iota)$, $\bG_0(\ii\tau_\kappa)$, and $\bR_{\HFB}$ from $\bU$, $\bV$, and $\beps$ 
		\State $\bR^{(0)} \gets \bR_{\HFB}$
		\State Compute $\bD^{(0)}$, $\bkappa^{(0)}$ from $\bR^{(0)}$
		\State $i \gets 0$
		\While{$\max{\abs{\bG^{(i)}(\ii\tau_\kappa)- \bG^{(i-1)}(\ii\tau_\kappa)}} > \varpi_1$}
			\State $i \gets i+1$
			\State Compute $\bchi_0(\ii\tau)$: $\forall \kappa, \ \bchi_0(\ii\tau_\kappa) \gets \bG^{(i-1)}(\ii\tau_\kappa) $ [see Eq.~\eqref{eq:xo_real}]
			\State Fourier transform $\bchi_0$: $\forall \iota, \ \bchi_0(\ii\omega_\iota) \gets \bchi_0(\ii\tau_\kappa)$
			\State Compute $\bchi(\ii\omega)$: $\forall \iota, \ {\bchi}(\ii\omega_\iota) \gets (\bI-\bchi_0(\ii\omega_\iota) \cdot \bV)^{-1} \cdot \bchi_0(\ii\omega_\iota)$
			\State Evaluate $E_\text{c}^{\GM}$ from ${\bchi}(\ii\omega_\iota)$ and ${\bchi_0}(\ii\omega_\iota))$ [see Eq.~\eqref{eq:GMc}] 
			\State Compute $\bW_p(\ii\omega)$: $\forall \iota, \ \bW_p(\ii\omega_\iota) \gets \bv \cdot {\bchi}(\ii\omega_\iota)\cdot\bv$ 
			\State Inverse Fourier transform $\bW_p$: $\forall \kappa, \ \bW_p(\ii\tau_\kappa)  \gets \bW_p(\ii\omega_\iota) $ 
			\State Compute $\bSig_\text{c}(\ii\tau)$: $\forall \kappa, \ \bSig_\text{c}(\ii\tau_\kappa) \gets ( \bG^{(i-1)}(\ii\tau_\kappa),\bW_p(\ii\tau_\kappa))$ 
            \State [see Eqs.~\eqref{eq:Sigma_he_c}--\eqref{eq:Sigma_eh_c}]
			\State Fourier transform $\bSig_\text{c}$: $\forall \iota, \ \bSig_\text{c}(\ii\omega_\iota) \gets \bSig_\text{c}(\ii\tau_\kappa) $ 
			\State $j \gets 0$
			\While{$\max{\abs{\bR^{(j)}- \bR^{(j-1)}}} > \varpi_2$}
				\State $j \gets j+1$
				\State Construct $\bH$ from $\bR^{(j-1)}$ and $\mu$
				\State Compute $\bG^{(i)}(\ii\omega)$: $\forall \iota, \ \bG^{(i)}(\ii\omega_\iota) \gets (\bH ,\bSig_\text{c}(\ii\omega_\iota))$ 
                \State [see Eq.~\eqref{eq:Dyson_w}]
				\State Evaluate $\bR^{(j)}$ from $\bG^{(i)}(\ii\omega_\iota)$ [see Eqs.~\eqref{eq:DR_from_DG} and \eqref{eq:R_fromG}]
				\State Compute $\bD^{(j)}$ and $\bkappa^{(j)}$ from  $\bR^{(j)}$
				\If{$ \abs{\Tr[ \bD^{(j)}]- N/2} > \varpi_2$} 
					\State Apply the bisection method to optimize $\mu$ in $\bH$
                    \State(This optimization also updates $\bG^{(i)}(\ii\omega_\iota)$ and $\bR^{(j)}$)
				\EndIf
				\State Apply DIIS to $\bR^{(j)}$ with error $\abs{\bR^{(j)} - \bR^{(j-1)}}$ 
			\EndWhile 
			\State Inverse Fourier transform $\bG^{(i)}$: $\forall \kappa, \ \bG^{(i)}(\ii\tau_\kappa) \gets \bG^{(i)}(\ii\omega_\iota)$ 
            \State [see Eq.~\eqref{eq:Ggorkov_from_itau_1}]
			\State Apply DIIS to $\bG^{(i)}(\ii\tau_\kappa)$ with error $\abs{\bG^{(i)}(\ii\tau_\kappa)- \bG^{(i-1)}(\ii\tau_\kappa)}$
		\EndWhile 
		\State Evaluate $E_{\HFB}$ using $\bD^{(j)}$ and $\bkappa^{(j)}$
		\State \Return $E_\HFB$, $E^{\GM}_\text{c}$, $\bG$, and $\bR$
	\EndProcedure
\end{algorithmic}
\caption{Bogoliubov self-consistent $GW$}
\label{alg:algo_bscgw}
\end{algorithm}

\section{Numerical examples}
\label{sec:results}
\begin{figure}
    \centering
    \includegraphics[width=0.62\linewidth]{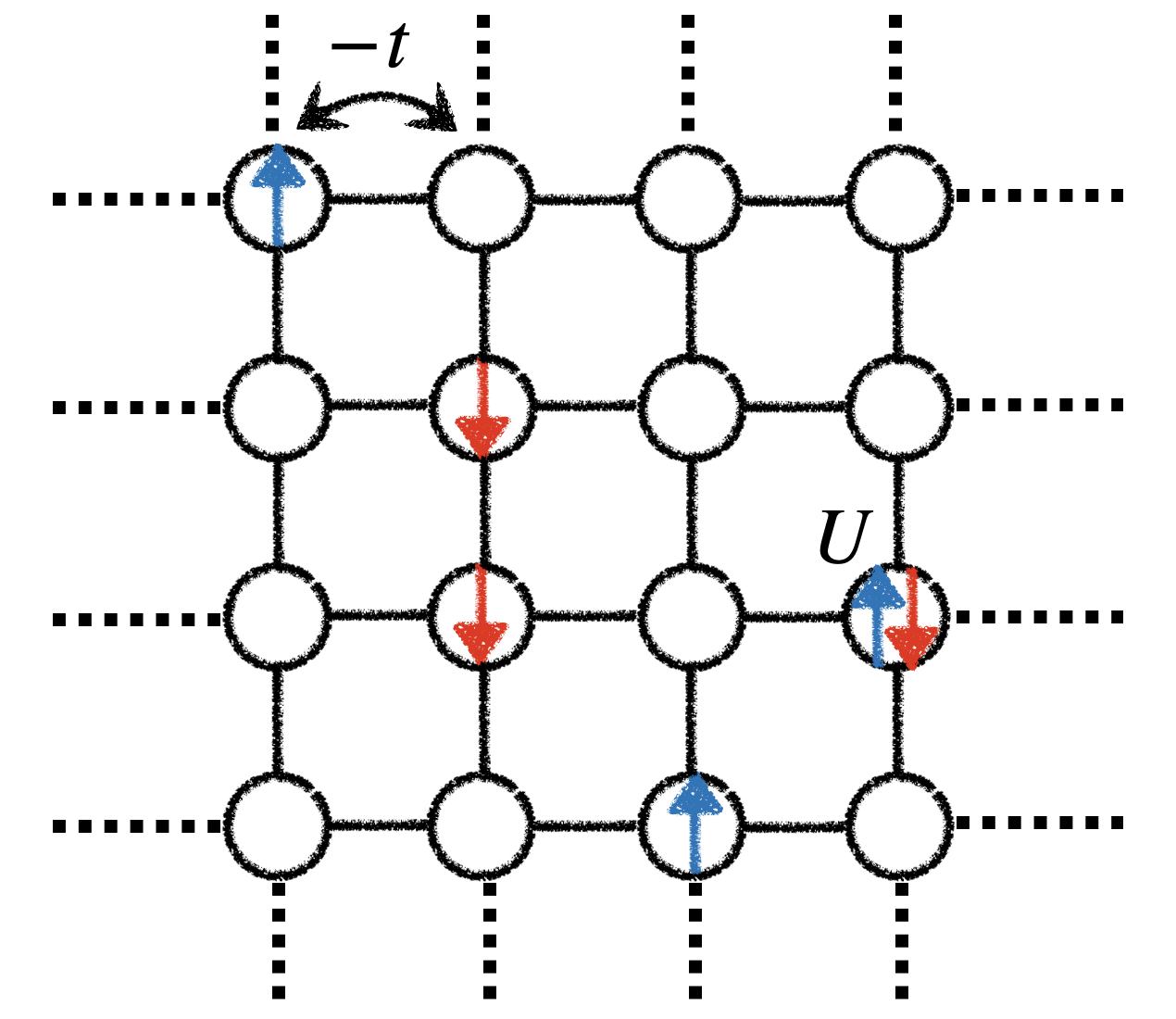}
    \caption{Schematic representation of the $4\times4$ Hubbard lattice with periodic boundary conditions (dotted lines). 
    The parameter $t$ denotes the nearest-neighbor hopping amplitude between lattice sites, while $U$ represents the on-site interaction between fermions of opposite spin (attractive in the present work). 
    Blue and red arrows depict spin-up and spin-down fermions, respectively.}
    \label{fig:4x4_lattice}
\end{figure}

\begin{figure}
    \centering
    \includegraphics[width=\linewidth]{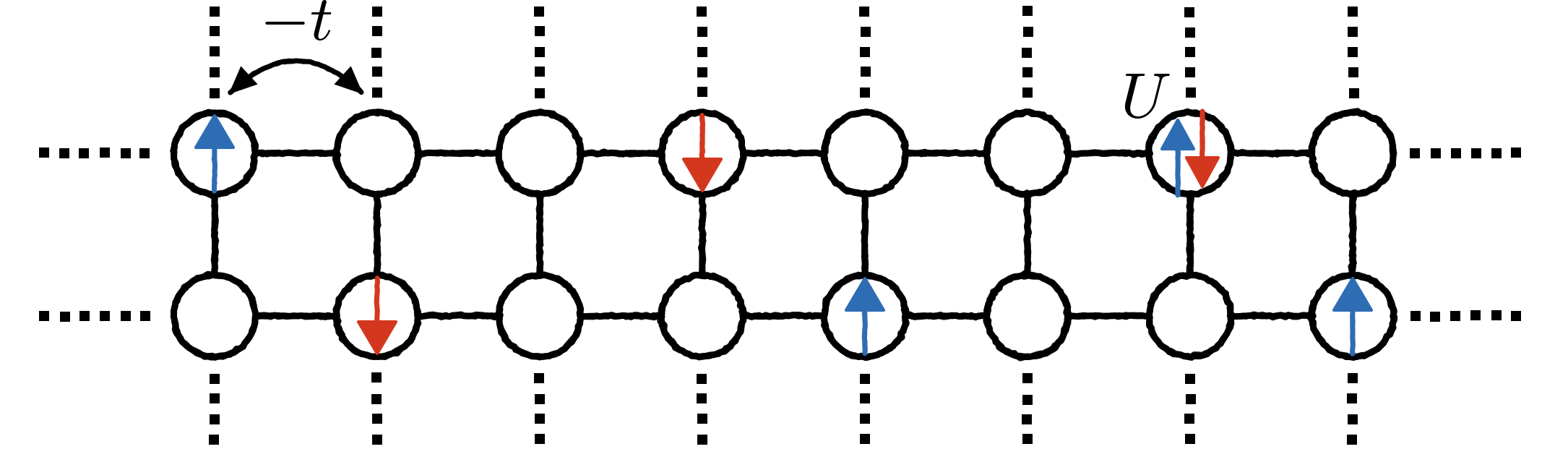}
    \caption{Schematic representation of the $2\times8$ Hubbard lattice with periodic boundary conditions (dotted lines). 
    Symbols are as in Fig.~\ref{fig:4x4_lattice}.}
    \label{fig:2x8_lattice}
\end{figure}

We assess the Bog.~sc$GW$ and scGF2 methods on the attractive Hubbard model, a minimal lattice model of fermions interacting through an on-site attraction. 
Despite its simplicity, it captures the essential physics of fermion pairing, including the crossover from BCS pairing of long-range Cooper pairs to the Bose--Einstein condensation of tightly bound pairs, which is relevant for superconductors, ultracold Fermi gases, and nuclear systems~\cite{micnas1990superconductivity,strinati2018bcs,hartke2023direct}. 
In the electronic context, it provides a minimal yet nontrivial description of a singlet $s$-wave superconductor in its ground state~\cite{lieb1989two,scalettar1989phase,moreo1991two,singer1998phase,goto2025superconductivity,karlsson2026}, while in nuclear physics its extended version has been used as a simple lattice model of nucleon matter treated at the HFB level~\cite{abe2004mean}. 
The model has also become a standard testbed for many-body methods designed for attractive interactions, including quantum Monte Carlo formulations that explicitly break $U(1)$ symmetry~\cite{shi2017many} and $GW$-based approaches~\cite{xiong2025application}. 
Being free of the specifics of any given material or nuclear force, it allows us to isolate the performance of the method itself for generic attractive fermionic interactions, and, for small clusters, FCI provides reference results. 
The Hamiltonian reads
\begin{equation}
    \hH=-t \sum_{\sigma=\downarrow,\uparrow}\sum_{\expval{\vartheta \nu}} ( \cre{a}{\vartheta \sigma}\anh{a}{\nu\sigma}  + \text{h.c.} ) +U \sum_{\nu} \hat{n}_{\nu\uparrow} \hat{n}_{\nu\downarrow}
\end{equation}   
where $\vartheta$ and $\nu$ label lattice sites (i.e., an orthonormal site basis), $\expval{\vartheta\nu}$ denotes nearest-neighbor pairs (including periodic boundary conditions), $t$ is the hopping amplitude, $\hat{n}_{\nu\sigma}=\cre{a}{\nu\sigma}\anh{a}{\nu\sigma}$ is the number operator, and $U<0$ is the attractive on-site interaction.
The two lattices considered in this work are depicted in Figs.~\ref{fig:4x4_lattice} and~\ref{fig:2x8_lattice}.

Because both $t$ and $U$ are real, the BCS ground-state wavefunction may be chosen real. Consequently, the eigenvectors of the Bogoliubov Hamiltonian $\bH$, and thus the matrices $\bU$ and $\bV$ entering the Bogoliubov transformation, can also be taken real. 
The normal and anomalous density matrices are therefore real as well and are naturally expressed in the site basis.

For this model, the interaction matrix elements simplify considerably. 
In the site basis, they reduce to
\begin{equation}
    v_{\gamma \zeta,\vartheta \nu} = U \delta_{\vartheta \nu} \delta_{\nu \gamma} \delta_{\gamma \zeta} 
\end{equation}
where $\delta_{\mu\nu}$ denotes the Kronecker delta. The independent-particle polarizability $\boldsymbol{\chi}_0$ is likewise expressed in the site basis, following Eq.~\eqref{eq:xo_real}.
\subsection{The 4$\times$4 and 2$\times$8 lattice Hubbard models}
\label{sec:4_4}

We consider two attractive Hubbard lattices with 16 sites and periodic boundary conditions: a $4\times4$ and a $2\times8$ lattice (see Figs.~\ref{fig:4x4_lattice} and~\ref{fig:2x8_lattice}). 
The $4\times4$ lattice has already been used for this purpose by Ref.~\cite{shi2017many}. 
The $2\times8$ lattice has the same number of sites but a different geometry, which allows us to probe the sensitivity of the results to the lattice topology. 

In Ref.~\cite{shi2017many}, calculations were performed in the strong-coupling regime for doped systems away from half-filling, so in this work we also target cases away from half-filling with total spin projection $S_z = 0$. \rev{Moreover, systems of interest for attractive interactions are usually found away from half-filling~\cite{scalettar1989phase,moreo1991two,micnas1990superconductivity}.} 
FCI results are available for these systems, providing the reference against which we benchmark the Bog.~sc$GW$ and scGF2 methods. 

In all calculations, we consider $N =10$ electrons in the $4\times4$ lattice and $N=6$ electrons in the $2\times8$ lattice. 
These fillings are chosen so that a closed-shell reference can be safely employed due to the large HOMO--LUMO gap that ensures that restricted HF (RHF) is able to describe these systems. 
This also facilitates our comparison of the Bogoliubov methods with their non-Bogoliubov counterparts (see Section I in the Supporting Information for more details).

In our implementation, the determination of the BCS state via the HFB method is preceded by a restricted HF calculation, as detailed in Appendix~\ref{app:hfb_algo} (see also Algorithm~\ref{alg:algo_hfb}). 
In practice, increasing the effective pairing strength leads to an enhancement of the anomalous density. 
The magnitude of the latter is quantified as $\norm{\bkappa}^2=\Tr[\bkappa^\dagger \cdot \bkappa]$, which we evaluate in the site basis. 

Next, after completing the HFB calculation, we proceed with the Bog.~sc$GW$ and scGF2 procedures, imposing, in all cases, the $\varpi_1 =\varpi_2 =  \num{d-6}$ thresholds. 
The imaginary-time and imaginary-frequency grids are constructed using $N_\tau = N_\omega = 30$ points, together with a regularization parameter of \num{d-5} through the \texttt{greenX} library. In principle, the parameters $e_\text{min}$ and $e_\text{max}$ required by the library, which define the frequency and time ranges, depend on the specific system. In standard $G_0W_0$ and qs$GW$ calculations, $e_\text{min}$ is typically set equal to the gap energy and $e_\text{max}$ to the largest occupied-virtual energy difference, which ensures accurate Fourier transforms of the polarizability~\cite{duchemin2021cubic,van2006quasiparticle}. However, in the fully self-consistent $GW$ framework, the Fourier transforms of the Green's function are also required, where the lowest exponential decay [see, for example, Eq.~\eqref{eq:G_he_t_im}] corresponds to that of the highest occupied single-particle state. For this reason, $e_\text{min}$ is more appropriately chosen as the energy of half the gap value in sc$GW$, after aligning the energy of the single-particle states with respect to the Fermi level. To cover all ranges of energies needed in this study, we have set $e_\text{min}=3.0$ and $e_\text{max}=20.0$ for all systems. With these grid settings, the numerical integration errors remain below \num{5d-3}. Based on our experience, this level of accuracy represents the minimum threshold required to obtain stable and internally consistent results. Finally, we emphasize that the same algorithm, including the adjustment of the chemical potential $\mu$ during the self-consistent cycle, was used for all the self-consistent Green's-function methods considered (sc$GW$, scGF2, and their Bogoliubov counterparts), and that this adjustment was essential to ensure numerical stability and to enforce the correct particle number. The Bog.\ CCSD calculations were performed using the same formalism as in Ref.~\cite{cui2025ab}, with the chemical potential subsequently updated based on the 1-RDM obtained from Bog.\ CCSD to reproduce the desired number of electrons. Bog.\ scGF2 calculations are based on the equations provided in the Supporting Information to build the self-energy, otherwise following the same algorithm as for Bog.~sc$GW$ (see Algorithm~\ref{alg:algo_bscgw}). \rev{As in other fully self-consistent Green's function methods~\cite{pokhilko2025homotopy}, multiple solutions may be found; for example, for Bog.~sc$GW$ in the $2\times8$ lattice, scanning $U/t$ in opposite directions leads to different solutions for $-2.8\lesssim U/t\lesssim-2.4$. To remain on a single solution, all calculations were performed as warm scans, using the converged solution at the previous value of $U/t$ as the initial guess. Tabulated total energies of all methods are provided in Section~V of the SI.}

\begin{figure}
    \centering
    \includegraphics[width=\linewidth]{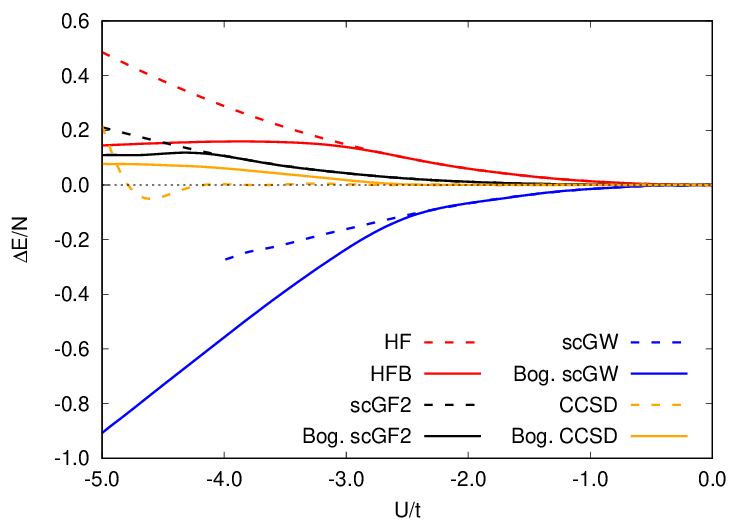}
    \caption{Error in the total energy per electron with respect to FCI, $\Delta E/N = (E - E_\text{FCI})/N$, versus $U/t$ for the $2\times8$ lattice with $N=6$. The horizontal line marks the FCI reference ($\Delta E/N = 0$).}
    \label{fig:DE_2x8}
\end{figure}

\begin{figure}
    \centering
    \includegraphics[width=\linewidth]{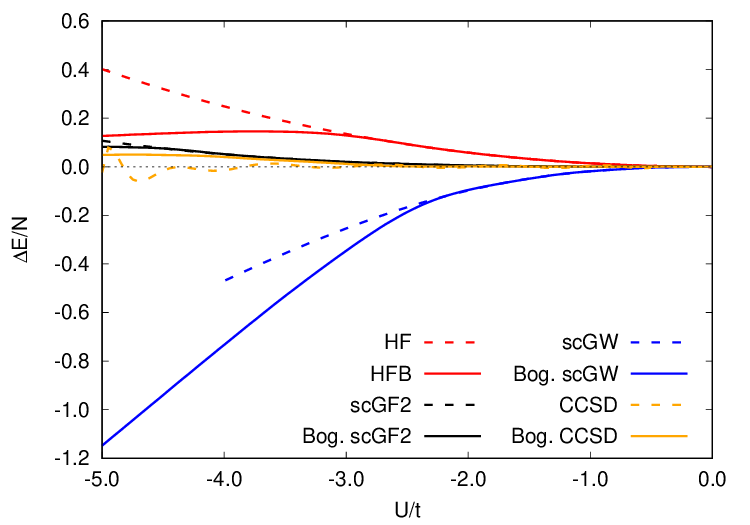}
    \caption{Error in the total energy per electron with respect to FCI, $\Delta E/N = (E - E_\text{FCI})/N$, versus $U/t$ for the $4\times4$ lattice with $N=10$. The horizontal line marks the FCI reference ($\Delta E/N = 0$).}
    \label{fig:DE_4x4}
\end{figure}

Figs.~\ref{fig:DE_2x8} and~\ref{fig:DE_4x4} show the error in the total energy per electron with respect to FCI, $\Delta E /N = (E - E_\text{FCI})/N$, as a function of $U/t$ for the $2\times8$ and $4\times4$ lattices, respectively. 
In addition to HF, HFB, and Bog.~sc$GW$, we add scGF2, sc$GW$, and CCSD, together with their Bogoliubov counterparts (Bog.\ scGF2, Bog.\ sc$GW$, and Bog.\ CCSD) for comparison. 
Among them, Bog.\ CCSD and Bog.\ scGF2 perform best in the range of $U/t$ values investigated, with the latter up to $\abs{U/t} = 5$. 

The curves of the two lattices are very similar despite their different topology and number of electrons; this suggests that neither the filling nor the geometry is the dominant factor. The HF error grows monotonically with the attraction (0.49 for the $2\times8$ lattice and 0.40 for the $4\times4$ lattice at $U/t=-5$), whereas HFB coincides with HF until pairing sets in at $U/t \approx -2.7$ and then lowers the error, which nevertheless remains sizable (0.14 for the $2\times8$ lattice and 0.13 for the $4\times4$ lattice at $U/t=-5$). 
Both sc$GW$ and Bog.\ sc$GW$ overestimate the correlation energy, yielding energies below FCI with errors that grow rapidly with the attraction and are larger for Bog.\ sc$GW$, up to $-0.91$ ($2\times8$) and $-1.15$ ($4\times4$) at $U/t=-5$ (sc$GW$ could only be converged up to $U/t \approx -4$).
The scGF2 and Bog.\ scGF2 energies are almost identical down to $U/t \approx -4$, and the Bogoliubov version improves only at stronger interactions, where its error grows to 0.11 for the $2\times8$ lattice and 0.08 for the $4\times4$ lattice at $U/t=-5$. 
Standard CCSD is very accurate at moderate attractions but becomes erratic and starts to diverge for $U/t \lesssim -4.5$, while Bog.\ CCSD is essentially exact for $U/t \gtrsim -2.5$ and its error grows smoothly up to about 0.08 for the $2\times8$ lattice and 0.05 for the $4\times4$ lattice at $U/t=-5$. Lastly, let us mention that for the systems employed, the Bog. scGF2 solution appears at higher $|U/t|$ values than the HFB one, indicating that scGF2 better accounts for correlation effects than HF (as expected). In contrast, the appearance of a Bog. sc$GW$ solution at lower $|U/t|$ values than the HFB one clearly shows that sc$GW$ misses important contributions, causing it to fail to describe the correlation effects present in attractive interactions. 

\begin{figure}
    \centering
    \includegraphics[width=\linewidth]{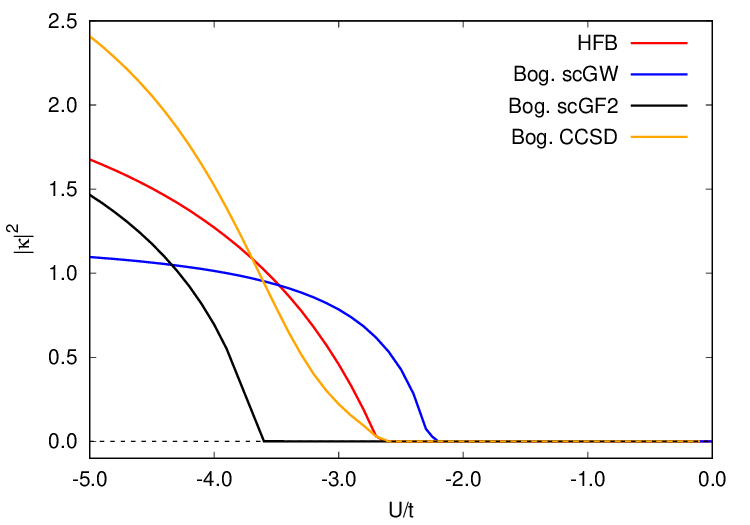}
    \caption{Magnitude of the anomalous density, $\norm{\bkappa}^2$, versus $U/t$ for the $2\times8$ lattice with $N=6$.}
    \label{fig:Norm_2x8}
\end{figure}

\begin{figure}
    \centering
    \includegraphics[width=\linewidth]{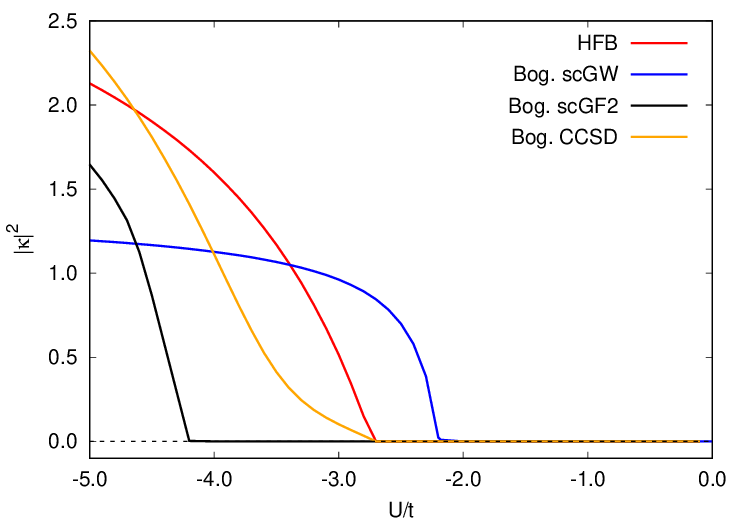}
    \caption{Magnitude of the anomalous density, $\norm{\bkappa}^2$, versus $U/t$ for the $4\times4$ lattice with $N=10$.}
    \label{fig:Norm_4x4}
\end{figure}

Figures~\ref{fig:Norm_2x8} and~\ref{fig:Norm_4x4} show the magnitude of the anomalous density.
It quantifies the degree of particle-number symmetry breaking introduced by the approximate methods, and thus the extent to which mixing different particle-number sectors is used to describe correlation effects.
Since the underlying Hamiltonian is particle-number conserving, the exact solution (FCI) does not exhibit any anomalous density.
In the case of the HFB solution, $\norm{\bkappa}^2$ grows steeply after $U/t \lesssim -2.7$, coinciding with the difference in energy between RHF and HFB seen in Figs.~\ref{fig:DE_2x8} and~\ref{fig:DE_4x4}. 
In Bog.\ sc$GW$ the onset of $\norm{\bkappa}^2$ is already at $U/t \approx -2.2$.
However, for $U/t \lesssim -3.4$ it yields a smaller $\norm{\bkappa}^2$ than HFB (about 1.1 and 1.2 at $U/t=-5$, i.e., a reduction of 35--45\%).
For Bog.\ scGF2 we find $\bkappa=\bO$ until $U/t \approx -3.9$ ($-4.3$) in the $2\times8$ ($4\times4$) lattice, consistent with its energies being almost identical to those of scGF2 in Figs.~\ref{fig:DE_2x8} and~\ref{fig:DE_4x4} up to that point, and then $\norm{\bkappa}^2$ grows rapidly. 
Finally, Bog.\ CCSD, yields the largest $\norm{\bkappa}^2$ at strong coupling (2.4 and 2.3 at $U/t=-5$).
Its onset coincides with HFB, since Bog.\ CCSD inherits the broken-symmetry reference from HFB; the symmetry of the reference mean-field state is preserved.
Whether this change in an orbital-optimized CCSD procedure is outside the scope of this work.

\section{Discussion}
\label{sec:discussions}
From the results presented in the previous section, it becomes apparent that sc$GW$ and Bog.~sc$GW$ resulted in a poor energetic description of the attractive Hubbard model. 
Note that similar findings have also been observed in Ref.~\cite{xiong2025application} for the attractive three-dimensional Hubbard model. 
We attribute this `failure' of sc$GW$ to possible instabilities in the particle-hole RPA, as presented in the SI (see Sec.~III).
While we investigate these instabilities for the RHF and HFB reference only, they might also persist for the RPA evaluated based on the renormalized propagators [see Eqs.~\eqref{eq:xo_ubar}-\eqref{eq:xo_hhee_2} and Eq.~\eqref{eq:Polarizability}].
This highlights the potential conceptual problem of sc$GW$ and Bog.~sc$GW$ in the presence of an effective attractive interaction, and is tightly bound to the self-screening error underlying this approximation~\cite{Nelson2007,Romaniello2009,Aryasetiawan2012,Wetherell2018}.  \rev{The overestimation of the correlation energy by sc$GW$ is not specific to attractive interactions. In molecular systems, studies have reported sc$GW$ total energies lying 0.10--0.11 a.u. below CCSD, whereas scGF2 energies lie 0.005--0.036 a.u. above it~\cite{pokhilko2024tensor}, and similar trends have been found for hydrogen chains~\cite{lan2017testing}. Our results follow the same pattern: for $U/t=-1$ to $-4$, sc$GW$ recovers 2.0--2.9 times the FCI correlation energy (defined with respect to RHF), whereas scGF2 recovers 63--96\%. To help interpret the part of the error that may be specific to the attractive case, we note that, at the RPA level, the Hessian analysis (see SI, Sec.~III) shows that, for attractive interactions, the RHF and HFB references may be unstable, unlike for repulsive interactions. This indicates that the RPA polarizability is likely a poor approximation for these systems, and it is the polarizability that ultimately determines the GM energy [Eq.~\eqref{eq:GMc}].}

In contrast, the self-interaction-free scGF2 method shows good performance throughout the investigated parameter range. 
To the best of our knowledge, Bog.~scGF2 is applied here for the first time to an electronic attractive model. 
The second-order Gorkov self-energy is, however, not new in nuclear physics. 
The fully self-consistent formalism of Ref.~\cite{soma2011ab} corresponds to our Bog.~scGF2, formulated in the energy domain for finite nuclei, and it is a cornerstone of \textit{ab initio} calculations of semimagic nuclei~\cite{soma2020self}. 
The Gorkov formalism was introduced there precisely because expansions around a non-superfluid reference break down in the presence of Cooper instabilities~\cite{soma2011ab}, which mirrors the deterioration of the standard (non-Bogoliubov) methods that we observe at strong attraction. 

\rev{We note that the anomalous density is a property of the symmetry-broken approximate state: the exact eigenstates of a particle-number-conserving Hamiltonian have $\bkappa=\bO$, so $\norm{\bkappa}^2$ in Figs.~\ref{fig:Norm_2x8} and~\ref{fig:Norm_4x4} measures the symmetry breaking introduced by each approximation and not a benchmarked quantity. In the fully self-consistent Green's function methods, correlations feed back into the propagator and delay or reduce this breaking with respect to HFB (the onset moves from $U/t\approx-2.7$ to $-3.9$ and $-4.3$ for Bog.~scGF2, and $\norm{\bkappa}^2$ is reduced for Bog.~sc$GW$), while only symmetry restoration removes it exactly~\cite{qiu2019particle}. A number-conserving definition of pairing in FCI, such as pair-pair correlation functions, goes beyond the scope of this work.}

Furthermore, the self-consistent Green's function implementation presented in this work allows for several extensions in terms of vertex corrections to improve upon sc$GW$~\cite{pokhilko2025tensor,kutepov2021electronic,wen2024comparing}, or alternative higher-order perturbation expansions already used in nuclear physics~\cite{soma2020self} to be incorporated. Access to total energies also enables response properties through energy gradients, such as electron-phonon couplings, which can be obtained from the correlated density matrices~\cite{pokhilko2021evaluation}, and calculations across a range of temperatures may give access to critical temperatures of pairing transitions.

\section{Conclusion}
\label{sec:conclusion}

In this work, we have evaluated the Bog.~sc$GW$ and scGF2 methods, implemented within an imaginary-time and imaginary-frequency framework to enable an efficient and stable optimization of the Gorkov propagator. 
The methods have been assessed against FCI results using the attractive Hubbard model. 
This model provides a minimal setting for pairing correlations between fermions, as found in superconductors~\cite{strinati2018bcs}, nuclei~\cite{abe2004mean}, and ultracold Fermi gases~\cite{hartke2023direct}, so our results offer \rev{only} a first indication of what can be expected from Bogoliubov self-consistent Green's function methods in these fields.

Our results show that Bog.~sc$GW$ performs poorly \rev{for the two 16-site lattices studied here}. 
It overestimates the correlation energy, yielding total energies well below the exact ones, with errors that grow rapidly with the attractive interaction and that, at strong coupling, are larger in magnitude than those of the HFB mean-field solution. 
Bog.~scGF2, obtained with the same algorithm by simply replacing the self-energy, remains accurate up to $\abs{U/t} \approx 4$, while Bog.~CCSD remains accurate over the whole range considered, which suggests the crucial role of exchange contributions, absent in the $GW$ self-energy, for these systems within many-body perturbation theory.

The surprisingly good accuracy of Bog.~scGF2 suggests that it may provide a useful, fully self-consistent Green's function framework for exploratory \textit{ab initio} studies of superconducting systems, at a formal cost lower than that of Bog.~CCSD, which has already been applied to electronic unconventional superconductors~\cite{cui2025ab}.
Nevertheless, as a second-order perturbative approach, its reliability for realistic strongly correlated systems should not be taken for granted.
We therefore view Bog.~scGF2 primarily as a potentially useful starting point for future investigations of more realistic superconducting and nuclear systems, rather than as a generally predictive method for such applications.

\rev{The two lattices considered here, despite their different geometry and number of electrons, show very similar trends, which suggests that these conclusions are not specific to a single case. Since this is a model study on two 16-site Hubbard lattices ($N=6$ and $N=10$) away from half-filling, studies on other fillings and larger systems could complement the present picture.}

Finally, although we focused on spin-unpolarized systems, the present formalism can be straightforwardly generalized to spin-polarized calculations and to the relativistic regime with two-component spinors~\cite{dyall2007introduction} (see, for example, Eq.~(A4) of Ref.~\cite{reho2024density}).
Spin-orbit coupling is particularly appealing, as it has been suggested to play a nontrivial role in certain unconventional superconductors~\cite{gastiasoro2020anisotropic}.

\begin{acknowledgments}
This project has received funding from the European Research Council (ERC) under the European Union's Horizon 2020 research and innovation program (Grant agreement no. 863481). 
J.~T.~acknowledges support from the Cluster of Excellence “CUI: Advanced Imaging of Matter” of the
Deutsche Forschungsgemeinschaft (DFG) (EXC 2056, funding ID 390715994, and funding by the Deutsche
Forschungsgemeinschaft (DFG, German Research Foundation)
- 577598009. 
The authors thank Carlos Mejuto-Zaera for fruitful discussions. 
M.~R-M also thanks Ivan Duchemin and Antoine Marie for the insightful discussions and helpful numerical comparisons. 
\end{acknowledgments}

\appendix

\section{The Hartree--Fock--Bogoliubov Algorithm}
\label{app:hfb_algo}

We begin the implementation by incorporating the HFB method for spin-unpolarized systems with real-valued single-particle states. 
The HFB calculation is initialized from a preceding spin-unpolarized (restricted) HF computation, which provides a set of canonical orbitals used as the starting point of the HFB procedure.
From these canonical orbitals, we generate fractional occupation numbers using a Fermi-Dirac distribution constructed from the HF orbital energies, $\eps_p^{\HF}$, a trial electronic temperature $T$, and a chemical potential $\mu$ adjusted to reproduce the target particle number $\expval{N}$
\begin{equation}
	n_p = \frac{1}{1+\exp\qty[\qty(\eps^{\HF}_p-\mu)/T]}
\end{equation}
This procedure provides a smooth initial guess for the occupation pattern and facilitates convergence of the subsequent HFB iterations.

The occupation numbers are first used to construct an initial guess for the normal density matrix, $\bD^{(0)}$, and, through Eq.~\eqref{eq:an_den_rel}, for the anomalous density, $\bkappa^{(0)}$. 
From these quantities, the initial HFB Hamiltonian, $\bH^{(0)}$, is built and diagonalized, yielding the quasiparticle eigenvectors $\bU^{(0)}$ and $\bV^{(0)}$. 
These eigenvectors are then employed to construct the generalized density matrix $\bR^{(1)}$, from which updated densities $\bD^{(1)}$ and $\bkappa^{(1)}$ are obtained for the next iteration.
Convergence is assessed through the commutator between the Hamiltonian and the generalized density matrix, using the criterion $\comm*{\bH^{(i)}}{\bR^{(i-1)}} < \varpi_3$, with $\varpi_3 = \num{d-8}$.
It is important to emphasize that the Hamiltonian $\bH$ depends explicitly on the chemical potential $\mu$. 
Consequently, $\mu$ must be readjusted whenever the updated eigenvectors $\bU^{(i)}$ and $\bV^{(i)}$ produce a density satisfying $\Tr[\bD^{(i)}] \neq N/2$. 
The chemical potential is efficiently optimized using the bisection method~\cite{burden1997numerical}, with a tolerance $\varpi_4$ of \num{d-8} on the deviation from the target value $N$.

A schematic overview of the complete HFB procedure is provided in Algorithm~\ref{alg:algo_hfb}.

\begin{algorithm}[H]
\begin{algorithmic}
\Procedure{HFB}{$\bD_{\HF}$}
	\State Determine $\bD^{(0)}$, $\bkappa^{(0)}$, and $\mu$ from Fermi distribution at $T$
	\State Compute $\bR^{(0)}$ and $\bH^{(0)}$ from $\bD^{(0)}$, $\bkappa^{(0)}$ and $\mu$
	\State $i \gets 0$
	\While{$ \max{ \abs{\comm{ \bH^{(i)} }{ \bR^{(i-1)} } }  }> \varpi_3$}
		\State $i \gets i+1$
		\State Diagonalize $\bH^{(i-1)}$ to get $\bU$, $\bV$, and $\beps$
		\State Compute $\bR^{(i)}$ from $\bU$ and $\bV$
		\State Form $\bD^{(i)}$ and $\bkappa^{(i)}$ from $\bR^{(i)}$
		\If{$\abs{\Tr[ \bD^{(i)}]- N/2)}>\varpi_4$}
			\State Apply the bisection method to optimize $\mu$ in $\bH^{(i-1)}$
            \State (This optimization also updates $\bR^{(i)}$)
		\EndIf
    \State Compute $\bH^{(i)}$ from $\bD^{(i)}$, $\bkappa^{(i)}$ and $\mu$ [see Eqs.~\eqref{eq:Fock} and~\eqref{eq:Delta}]
	\EndWhile
	\State \Return $\bV$, $\bU$, $\beps$, and $\mu$
\EndProcedure
\end{algorithmic}
\caption{Self-consistent Hartree--Fock--Bogoliubov}
\label{alg:algo_hfb}
\end{algorithm}

\section{Quadrature Grids}
\label{app:grids}
As discussed in Sec.~\ref{subsec:bogoliubov_gw}, the Gorkov propagator, the polarizability, and the self-energy can be represented as matrices either in imaginary time or in imaginary frequency. 
When expressed in imaginary time, the Gorkov propagator yields the generalized density matrix in the limit $\tau \to 0^{-}$ [see Eq.~\eqref{eq:G_he_t_im}]. 
However, $\bR$ can also be obtained directly from the imaginary-frequency representation of the propagator via the integral~\cite{kaltak2014low}
\begin{equation}
    \bR = \frac{1}{2\pi}  \int ^{\infty} _{-\infty} d{\omega} \; \bG (\ii\omega)
\end{equation}

Similarly, the evaluation of the Galitskii–Migdal correlation energy [see Eq.~\eqref{eq:GMc}] requires an integration along the imaginary-frequency axis. 
To perform such integrals efficiently, various tailored numerical grids have been developed~\cite{kaltak2014low,grumet2018beyond,duchemin2020robust,duchemin2021cubic,duchemin2019separable,forster2021low,forster2023two,wilhelm2018toward,wilhelm2021low,azizi2023time,azizi2024validation}. 
Using these grids, the generalized density matrix can be approximated as
\begin{equation}
    \bR \approx \frac{1}{\pi}\sum ^{N_\omega} _\iota \eta_\iota \Re \left[ \bG (\ii\omega_\iota)\right]
        \label{eq:R_from_G}
\end{equation}
and the Galitskii--Migdal correlation energy as
\begin{equation}
    E_\text{c}^{\GM}[\bG] \approx \frac{1}{2\pi} \sum^{N_\omega} _\iota  \eta_\iota 
   \Tr \qty{ \bv \cdot \qty[ \bchi_0(\ii \omega_\iota) -  \bchi(\ii \omega_\iota) ] }
\end{equation}
where $\omega_\iota$ and $\eta_\iota$ denote the quadrature points and weights, respectively. 
Importantly, these schemes achieve high accuracy with a relatively small number of frequency points $N_\omega$.

In contrast, the construction of $\bchi_0(\ii\tau)$ and $\bSig_\text{c}(\ii\tau)$ requires a grid of imaginary times $\{\tau_\kappa\}_{\kappa\in[1,N_\tau]}$. 
Consequently, consistent and accurate transformations between imaginary-time and imaginary-frequency grids are essential. 
These transformations can be performed efficiently by decomposing the Fourier integrals into cosine and sine components and computing the corresponding quadrature weights using predefined target functions~\cite{kaltak2014low,grumet2018beyond,azizi2023time,azizi2024validation}.

In this work, we employ the recently developed \texttt{greenX} library~\cite{azizi2023time,azizi2024validation} to generate the imaginary-time and imaginary-frequency grids, as well as the associated transformation weights. 
In addition, we have implemented a missing sine transformation required to convert quantities from imaginary frequency to imaginary time. 
This development is now available in the official repository (\url{https://github.com/nomad-coe/greenX.git}).

As an illustration, the Fourier transform of the self-energy reads
\begin{equation}
\begin{split}
    \bSig_\text{c}(\ii\omega_\iota) &= -\ii \int ^{\infty} _{-\infty}d\tau \;  \bSig_c(\ii\tau) \exp(-\ii\omega_\iota \tau) 
    \\
    & = 2\int^{\infty}_0 \dd{\tau}  \bSig_\text{c}^{\cos}(\ii\tau) \cos(\omega_\iota\tau)  
    \\
    & +2 \int^{\infty}_0 \dd{\tau} \bSig_\text{c}^{\sin}(\ii\tau)  \sin(\omega_\iota\tau)  
    \\
    &\approx \sum^{N_\tau}_\kappa \eta^{\cos}_{\kappa,\iota} \bSig_\text{c}^{\cos}(\ii\tau_\kappa)   
    + \sum^{N_\tau}_\kappa \eta^{\sin}_{\kappa,\iota} \bSig_\text{c}^{\sin}(\ii\tau_\kappa)   
\end{split}
\end{equation}
where $\eta^{\cos}_{\kappa,\iota}$ and $\eta^{\sin}_{\kappa,\iota}$ are the quadrature weights provided by \texttt{greenX} for performing the transformation from the imaginary-time grid ${\tau_\kappa}$ to the frequency point $\omega_\iota$.
The auxiliary functions entering the cosine and sine transforms are defined as
\begin{subequations}
\begin{align}
    \bSig_\text{c}^{\cos}(\ii\tau) & = -\frac{\ii}{2} \qty[  \bSig_\text{c}(\ii\tau) +  \bSig_\text{c}(-\ii\tau) ]
    \\
    \bSig_\text{c}^{\sin}(\ii\tau) & = -\frac{1}{2} \qty[  \bSig_\text{c}(\ii\tau) -  \bSig_\text{c}(-\ii\tau) ]
\end{align}
\end{subequations}
which enforce the appropriate symmetry properties required by the cosine (even) and sine (odd) components of the transformation.

Finally, we recall that $\bchi_0$ and $\bW_p$ are even functions of both $\tau$ and $\omega$; therefore, their (inverse) Fourier transforms involve only cosine components. In contrast, the Gorkov propagator $\bG(\ii\omega)$ is neither purely even nor purely odd, and its transformation to imaginary time requires both sine and cosine contributions to recover $\bG(\pm \ii\tau_\kappa)$. 

\section{$\bG_0$ in Frequency Space}
\label{app:gorkov_freq}
The equations below are taken from Ref.~\cite{soma2011ab} and give the explicit expressions of the uncorrelated Gorkov propagator
\begin{subequations}
\begin{align}
	[\bG_0^{\he}(\omega)]^{\upar\upar}_{pq} 
	& =  \sum_I \left[ \frac{U_{pI}^\upar \qty(U_{qI}^\upar)^{*}}{\omega-\epsilon_I+\ii\eta}  +  \frac{\qty(V_{pI}^\upar)^{*} V_{qI}^\upar}{\omega+\epsilon_I-\ii\eta} \right]
 \label{eq:G_he_w}
\\
	[\bG_0^{\hh}(\omega)]^{\upar\dwar}_{pq} 
	& = \sum_I \left[ \frac{U_{pI}^\upar \qty(V_{qI}^\dwar)^{*}}{\omega-\epsilon_I+\ii\eta} +  \frac{\qty(V_{pI}^\upar)^{*} U_{qI}^\dwar}{\omega+\epsilon_I-\ii\eta}  \right]                      
\label{eq:G_hh_w}
\\
	[\bG_0^{\ee}(\omega)]^{\dwar\upar}_{pq}
	& = \sum_I \left[ \frac{V_{pI}^\dwar \qty(U_{qI}^\upar)^{*}}{\omega-\epsilon_I+\ii\eta}  + \frac{\qty(U_{pI}^\dwar)^{*} V_{qI}^\upar}{\omega+\epsilon_I-\ii\eta}  \right]
\label{eq:G_ee_w}
\\
	[\bG_0^{\eh}(\omega)]^{\dwar\dwar}_{pq} 
	& = \sum_I \left[ \frac{V_{pI}^\dwar \qty(V_{qI}^\dwar)^{*}}{\omega-\epsilon_I+\ii\eta}  +  \frac{\qty(U_{pI}^\dwar)^{*} U_{qI}^\dwar}{\omega+\epsilon_I-\ii\eta} \right]
\label{eq:G_eh_w}
\end{align}
\end{subequations}
with $\eta$ being a positive infinitesimal number, which can be set to zero when imaginary frequencies ($\ii\omega$) are employed.


\bibliography{my_biblio}

\end{document}